# The JCMT BISTRO Survey: An 850/450 μm Polarization Study of NGC 2071IR in Orion B

A-Ran Lyo[1], Jongsoo Kim[1,2], Sarah Sadavoy[3], Doug Johnstone[4,5], David Berry[6], Kate Pattle[7], Woojin Kwon[8,9], Pierre Bastien[10], Takashi Onaka[11,12], James Di Francesco[4,5], Ji-Hyun Kang[1], Ray Furuya[13,14], Charles L. H. Hull[15,16,85], Motohide Tamura[11,17,18], Patrick M. Koch[19], Derek Ward-Thompson[20], Tetsuo Hasegawa[17], Thiem Hoang[1,2], Doris Arzoumanian[21,22], Chang Won Lee[1,2], Chin-Fei Lee[19], Do-Young Byun[1], Florian Kirchschlager[23], Yasuo Doi[24], Kee-Tae Kim[1,2], Jihye Hwang[1,2], Pham Ngoc Diep[25], Lapo Fanciullo[19], Sang-Sung Lee[1,2], Geumsook Park[1], Hyunju Yoo[1], Eun Jung Chung[26], Anthony Whitworth[27], Steve Mairs[6], Archana Soam[28], Tie Liu[29], Xindi Tang[30], Simon Coudé[28], Philippe André[31], Tyler L. Bourke[32,33], Huei-Ru Vivien Chen[19,34], Zhiwei Chen[35], Wen Ping Chen[36], Mike Chen[5], Tao-Chung Ching[37,38], Jungyeon Cho[26], Minho Choi[1], Yunhee Choi[1], Antonio Chrysostomou[32], Sophia Dai[38], C. Darren Dowell[39], Hao-Yuan Duan[34], Yan Duan[38], David Eden[40], Chakali Eswaraiah[37,38], Stewart Eyres[41], Jason Fiege[42], Laura M. Fissel[3], Erica Franzmann[42], Per Friberg[6], Rachel Friesen[43], Gary Fuller[33], Tim Gledhill[44], Sarah Graves[6], Jane Greaves[27], Matt Griffin[27], Qilao Gu[45], Ilseung Han[1,2], Jannifer Hatchell[46], Saeko Hayashi[47], Martin Houde[48], Tsuyoshi Inoue[49], Shu-ichiro Inutsuka[49], Kazunari Iwasaki[50], Il-Gyo Jeong[51], Miju Kang[1], Akimasa Kataoka[52], Koji Kawabata[53,54,55], Francisca Kemper[19,56], Gwanjeong Kim[57], Mi-Ryang Kim[1], Shinyoung Kim[1,2], Kyoung Hee Kim[1], Jason Kirk[20], Masato I. N. Kobayashi[58], Vera Könyves[20], Takayoshi Kusune[17], Jungmi Kwon[11], Kevin Lacaille[59,60], Shih-Ping Lai[19,34], Chi-Yan Law[45,61], Jeong-Eun Lee[62], Yong-Hee Lee[62], Hyeseung Lee[1], Dalei Li[63], Di Li[37], Hua-Bai Li[45], Hong-Li Liu[64], Junhao Liu[65,66], Sheng-Yuan Liu[19], Xing Lu[17], Masafumi Matsumura[67], Brenda Matthews[4,5], Gerald Moriarty-Schieven[4], Tetsuya Nagata[68], Fumitaka Nakamura[69,70], Hiroyuki Nakanishi[71], Nguyen Bich Ngoc[25], Nagayoshi Ohashi[19], Harriet Parsons[6], Nicolas Peretto[27], Felix Priestley[27], Tae-soo Pyo[47,70], Lei Qian[37], Keping Qiu[65,66], Ramprasad Rao[19], Jonathan Rawlings[23], Mark G. Rawlings[6], Brendan Retter[27], John Richer[72,73], Andrew Rigby[27], Hiro Saito[74], Giorgio Savini[75], Anna Scaife[33], Masumichi Seta[76], Yoshito Shimajiri[77], Hiroko Shinnaga[71], Mehrnoosh Tahani[78], Ya-Wen Tang[19], Kohji Tomisaka[52,70], Le Ngoc Tram[79], Yusuke Tsukamoto[71], Serena Viti[80], Jia-Wei Wang[19], Hongchi Wang[35], Jinjin Xie[38], Hsi-Wei Yen[19], Jinghua Yuan[38], Hyeong-Sik Yun[62], Tetsuya Zenko[68], Guoyin Zhang[37], Chuan-Peng Zhang[37,38], Yapeng Zhang[81], Jianjun Zhou[63], Lei Zhu[37], Ilse de Looze[80], C. Darren Dowell[39], Sam Falle[82], Jean-François Robitaille[83], and Sven van Loo[84]

[1] Korea Astronomy and Space Science Institute, 776 Daedeokdae-ro, Yuseong-gu, Daejeon 34055, Republic of Korea
[2] University of Science and Technology, Korea, 217 Gajeong-ro, Yuseong-gu, Daejeon 34113, Republic of Korea
[3] Department for Physics, Engineering Physics and Astrophysics, Queen's University, Kingston, ON, K7L 3N6, Canada
[4] NRC Herzberg Astronomy and Astrophysics, 5071 West Saanich Road, Victoria, BC V9E 2E7, Canada
[5] Department of Physics and Astronomy, University of Victoria, Victoria, BC V8W 2Y2, Canada
[6] East Asian Observatory, 660 N. A'ohōkū Place, University Park, Hilo, HI 96720, USA
[7] Centre for Astronomy, School of Physics, National University of Ireland Galway, University Road, Galway, Ireland
[8] Department of Earth Science Education, Seoul National University, 1 Gwanak-ro, Gwanak-gu, Seoul 08826, Republic of Korea
[9] SNU Astronomy Research Center, Seoul National University, 1 Gwanak-ro, Gwanak-gu, Seoul 08826, Republic of Korea
[10] Centre de recherche en astrophysique du Québec & département de physique, Université de Montréal, 1375, Avenue Thérèse-Lavoie-Roux, Montréal, QC, H2V OB3, Canada
[11] Department of Astronomy, Graduate School of Science, The University of Tokyo, 7-3-1 Hongo, Bunkyo-ku, Tokyo 113-0033, Japan
[12] Department of Physics, Faculty of Science and Engineering, Meisei University, 2-1-1 Hodokubo, Hino, Tokyo 1191-8506, Japan
[13] Tokushima University, Minami Jousanajima-machi 1-1, Tokushima 770-8502, Japan
[14] Institute of Liberal Arts and Sciences Tokushima University, Minami Jousanajima-machi 1-1, Tokushima 770-8502, Japan
[15] National Astronomical Observatory of Japan, NAOJ Chile, Alonso de Córdova 3788, Office 61B, 7630422, Vitacura, Santiago, Chile
[16] Joint ALMA Observatory, Alonso de Córdova 3107, Vitacura, Santiago, Chile
[17] National Astronomical Observatory of Japan, National Institute of Natural Sciences, Osawa, Mitaka, Tokyo 181-8588, Japan
[18] Astrobiology Center, National Institutes of Natural Sciences, 2-21-1 Osawa, Mitaka, Tokyo 181-8588, Japan
[19] Academia Sinica Institute of Astronomy and Astrophysics, No.1, Sec. 4., Roosevelt Road, Taipei 10617, Taiwan
[20] Jeremiah Horrocks Institute, University of Central Lancashire, Preston PR1 2HE, UK
[21] Instituto de Astrofísica e Ciências do Espaço, Universidade do Porto, CAUP, Rua das Estrelas, PT4150-762 Porto, Portugal
[22] Department of Physics, Graduate School of Science, Nagoya University, Furo-cho, Chikusa-ku, Nagoya 464-8602, Japan
[23] Department of Physics and Astronomy, University College London, WC1E 6BT London, UK
[24] Department of Earth Science and Astronomy, Graduate School of Arts and Sciences, The University of Tokyo, 3-8-1 Komaba, Meguro, Tokyo 153-8902, Japan
[25] Vietnam National Space Center, Vietnam Academy of Science and Technology, 18 Hoang Quoc Viet, Hanoi, Vietnam
[26] Department of Astronomy and Space Science, Chungnam National University, 99 Daehak-ro, Yuseong-gu, Daejeon 34134, Republic of Korea
[27] School of Physics and Astronomy, Cardiff University, The Parade, Cardiff, CF24 3AA, UK
[28] SOFIA Science Center, Universities Space Research Association, NASA Ames Research Center, Moffett Field, CA 94035, USA
[29] Key Laboratory for Research in Galaxies and Cosmology, Shanghai Astronomical Observatory, Chinese Academy of Sciences, 80 Nandan Road, Shanghai 200030, People's Republic of China
[30] Xinjiang Astronomical Observatory, Chinese Academy of Sciences, 830011 Urumqi, People's Republic of China
[31] Laboratoire AIM CEA/DSM-CNRS-Université Paris Diderot, IRFU/Service d'Astrophysique, CEA Saclay, F-91191 Gif-sur-Yvette, France
[32] SKA Organisation, Jodrell Bank, Lower Withington, Macclesfield, SK11 9FT, UK
[33] Jodrell Bank Centre for Astrophysics, School of Physics and Astronomy, University of Manchester, Oxford Road, Manchester, M13 9PL, UK
[34] Institute of Astronomy and Department of Physics, National Tsing Hua University, Hsinchu 30013, Taiwan

[35] Purple Mountain Observatory, Chinese Academy of Sciences, 2 West Beijing Road, 210008 Nanjing, People's Republic of China
[36] Institute of Astronomy, National Central University, Zhongli 32001, Taiwan
[37] CAS Key Laboratory of FAST, National Astronomical Observatories, Chinese Academy of Sciences, People's Republic of China
[38] National Astronomical Observatories, Chinese Academy of Sciences, A20 Datun Road, Chaoyang District, Beijing 100012, People's Republic of China
[39] Jet Propulsion Laboratory, M/S 169-506, 4800 Oak Grove Drive, Pasadena, CA 91109, USA
[40] Astrophysics Research Institute, Liverpool John Moores University, IC2, Liverpool Science Park, 146 Brownlow Hill, Liverpool, L3 5RF, UK
[41] University of South Wales, Pontypridd, CF37 1DL, UK
[42] Department of Physics and Astronomy, The University of Manitoba, Winnipeg, Manitoba R3T2N2, Canada
[43] National Radio Astronomy Observatory, 520 Edgemont Road, Charlottesville, VA 22903, USA
[44] School of Physics, Astronomy & Mathematics, University of Hertfordshire, College Lane, Hatfield, Hertfordshire AL10 9AB, UK
[45] Department of Physics, The Chinese University of Hong Kong, Shatin, N.T., Hong Kong
[46] Physics and Astronomy, University of Exeter, Stocker Road, Exeter EX4 4QL, UK
[47] Subaru Telescope, National Astronomical Observatory of Japan, 650 N. A'ohōkū Place, Hilo, HI 96720, USA
[48] Department of Physics and Astronomy, The University of Western Ontario, 1151 Richmond Street, London N6A 3K7, Canada
[49] Department of Physics, Graduate School of Science, Nagoya University, Furo-cho, Chikusa-ku, Nagoya 464-8602, Japan
[50] Department of Environmental Systems Science, Doshisha University, Tatara, Miyakodani 1-3, Kyotanabe, Kyoto 610-0394, Japan
[51] Department of Astronomy and Atmospheric Sciences, College of Natural Sciences, Kyungpook National University, 80 Daehakro, Bukgu, Daegu 41566, Republic of Korea
[52] Division of Theoretical Astronomy, National Astronomical Observatory of Japan, National Institute of Natural Sciences, Osawa, Mitaka, Tokyo 181-8588, Japan
[53] Hiroshima Astrophysical Science Center, Hiroshima University, Kagamiyama 1-3-1, Higashi-Hiroshima, Hiroshima 739-8526, Japan
[54] Department of Physics, Hiroshima University, Kagamiyama 1-3-1, Higashi-Hiroshima, Hiroshima 739-8526, Japan
[55] Core Research for Energetic Universe (CORE-U), Hiroshima University, Kagamiyama 1-3-1, Higashi-Hiroshima, Hiroshima 739-8526, Japan
[56] European Southern Observatory, Karl-Schwarzschild-Str. 2, D-85748 Garching, Germany
[57] Nobeyama Radio Observatory, National Astronomical Observatory of Japan, National Institutes of Natural Sciences, Nobeyama, Minamimaki, Minamisaku, Nagano 384-1305, Japan
[58] Astronomical Institute, Graduate School of Science, Tohoku University, Aoba-ku, Sendai, Miyagi 980-8578, Japan
[59] Department of Physics and Astronomy, McMaster University, Hamilton, ON L8S 4M1 Canada
[60] Department of Physics and Atmospheric Science, Dalhousie University, Halifax B3H 4R2, Canada
[61] Department of Space, Earth & Environment, Chalmers University of Technology, SE-412 96 Gothenburg, Sweden
[62] School of Space Research, Kyung Hee University, 1732 Deogyeong-daero, Giheung-gu, Yongin-si, Gyeonggi-do 17104, Republic of Korea
[63] Xinjiang Astronomical Observatory, Chinese Academy of Sciences, 150 Science 1-Street, 830011 Urumqi, Xinjiang, People's Republic of China
[64] Department of Astronomy, Yunnan University, Kunming, 650091, People's Republic of China
[65] School of Astronomy and Space Science, Nanjing University, 163 Xianlin Avenue, Nanjing 210023, People's Republic of China
[66] Key Laboratory of Modern Astronomy and Astrophysics (Nanjing University), Ministry of Education, Nanjing 210023, People's Republic of China
[67] Faculty of Education & Center for Educational Development and Support, Kagawa University, Saiwai-cho 1-1, Takamatsu, Kagawa, 760-8522, Japan
[68] Department of Astronomy, Graduate School of Science, Kyoto University, Sakyo-ku, Kyoto 606-8502, Japan
[69] Division of Theoretical Astronomy, National Astronomical Observatory of Japan, Mitaka, Tokyo 181-8588, Japan
[70] SOKENDAI (The Graduate University for Advanced Studies), Hayama, Kanagawa 240-0193, Japan
[71] Department of Physics and Astronomy, Graduate School of Science and Engineering, Kagoshima University, 1-21-35 Korimoto, Kagoshima, Kagoshima 890-0065, Japan
[72] Astrophysics Group, Cavendish Laboratory, J.J. Thomson Avenue, Cambridge CB3 0HE, UK
[73] Kavli Institute for Cosmology, Institute of Astronomy, University of Cambridge, Madingley Road, Cambridge, CB3 0HA, UK
[74] Faculty of Pure and Applied Sciences, University of Tsukuba, 1-1-1 Tennodai, Tsukuba, Ibaraki 305-8577, Japan
[75] OSL, Physics & Astronomy Dept., University College London, WC1E 6BT London, UK
[76] Department of Physics, School of Science and Technology, Kwansei Gakuin University, 2-1 Gakuen, Sanda, Hyogo 669-1337, Japan
[77] National Astronomical Observatory of Japan, National Institutes of Natural Sciences, Osawa, Mitaka, Tokyo 181-8588, Japan
[78] Dominion Radio Astrophysical Observatory, Herzberg Astronomy and Astrophysics Research Centre, National Research Council Canada, P.O. Box 248, Penticton, BC V2A 6J9 Canada
[79] University of Science and Technology of Hanoi, Vietnam Academy of Science and Technology, 18 Hoang Quoc Viet, Hanoi, Vietnam
[80] Physics & Astronomy Dept., University College London, WC1E 6BT London, UK
[81] Department of Astronomy, Beijing Normal University, Beijing100875, People's Republic of China
[82] Department of Applied Mathematics, University of Leeds, Woodhouse Lane, Leeds LS2 9JT, UK
[83] Univ. Grenoble Alpes, CNRS, IPAG, F-38000 Grenoble, France
[84] School of Physics and Astronomy, University of Leeds, Woodhouse Lane, Leeds LS2 9JT, UK

## Abstract

We present the results of simultaneous 450 $\mu$m and 850 $\mu$m polarization observations toward the massive star-forming region NGC 2071IR, a target of the BISTRO (*B*-fields in STar-forming Region Observations) Survey, using the POL-2 polarimeter and SCUBA-2 camera mounted on the James Clerk Maxwell Telescope. We find a pinched magnetic field morphology in the central dense core region, which could be due to a rotating toroidal disklike structure and a bipolar outflow originating from the central young stellar object IRS 3. Using the modified Davis–Chandrasekhar–Fermi method, we obtain a plane-of-sky magnetic field strength of $563 \pm 421~\mu$G in the central $\sim 0.12$ pc region from 850 $\mu$m polarization data. The corresponding magnetic energy density of $2.04 \times 10^{-8}~\mathrm{erg~cm^{-3}}$ is comparable to the turbulent and gravitational energy densities in the region. We find that the magnetic field direction is very well aligned with the whole of the IRS 3 bipolar outflow structure. We find that the median value of polarization fractions is 3.0% at 450 $\mu$m in the central 3′ region, which is larger than the median value of 1.2% at 850 $\mu$m. The trend could be due to the better alignment of warmer dust in the strong radiation environment. We also find that polarization fractions decrease with intensity at both wavelengths, with slopes, determined by fitting a Rician noise model of $0.59 \pm 0.03$ at 450 $\mu$m and $0.36 \pm 0.04$ at 850 $\mu$m, respectively. We think that the shallow slope at 850 $\mu$m is due to grain alignment at the center being assisted by strong radiation from the central young stellar objects.

---

[85] NAOJ Fellow.

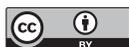

*Unified Astronomy Thesaurus concepts:* Star forming regions (1565); Interstellar magnetic fields (845); Interstellar medium (847); Polarimetry (1278)

## 1. Introduction

Magnetic fields are expected to influence the process of star formation, but their exact role at each evolutionary stage is not yet understood. The well-ordered magnetic field structures, which are observed from molecular cloud scales to protostellar scales, suggest that magnetic fields are an important part of the star formation process (e.g., Orion molecular cloud, Li et al. 2009; Taurus molecular cloud, Chapman et al. 2011; Orion A region, Pattle et al. 2018; NGC 1333 IRAS4A, Girart et al. 2006, etc.). Theoretical studies predict that magnetic fields affect core collapse, disk, and outflow formation, multiplicity of protostars, and the star formation rate (Shu et al. 1987, 2004; Allen et al. 2003; Nakamura & Li 2005; Price & Bate 2007; Kudoh & Basu 2008; Machida et al. 2011, 2014).

Dust polarization observations at submillimeter wavelengths are a good way to study dust properties as well as magnetic field properties in star-forming regions, most of which are heavily embedded in dense molecular clouds (Davis 1951; Goodman et al. 1995; Lazarian 2003). The observed polarization signature is attributed to aspherical dust grains that have their long axes aligned perpendicular to the local magnetic field direction (see reviews, e.g., Andersson et al. 2015; Lazarian et al. 2015; Pattle & Fissel 2019). Aligned dust grains result in the polarization of background starlight at optical/NIR wavelengths (Hall 1949; Hiltner 1949) and polarized thermal dust emission at far-IR/submillimeter wavelengths (Hildebrand 1989). The BISTRO (B-fields In STar-forming Region Observations) Survey project (Ward-Thompson et al. 2017), using the POL-2 polarimeter (Bastien et al. 2005a, 2005b; Friberg et al. 2016) on the Submillimetre Common-User Bolometer Array 2 (SCUBA-2) camera (Holland et al. 2013) mounted on the James Clerk Maxwell Telescope (JCMT), provides dust polarization maps toward nearby star-forming regions at 450 $\mu$m and 850 $\mu$m with angular resolutions of 9″6 and 14″1, respectively (Dempsey et al. 2013). This project provides opportunities for the study of dust properties and magnetic field morphology around star-forming regions at linear scales of 2000 ∼ 5500 au. These observations therefore well fill the gap between the few-pc-scale magnetic field structures revealed by optical/infrared observations (e.g., Tamura et al. 2007; Li et al. 2009; Chapman et al. 2011) and the few-au-scale magnetic field structures seen in submillimeter/millimeter interferometry observations (e.g., Girart et al. 2006; Kwon et al. 2019).

Comparison of the magnetic field strength and structure with outflow properties is important in order to understand the star formation process. On the relatively large dense core scale of ∼0.1 pc, the rotational axis of a core (the outflow direction) is theoretically expected to be aligned with the magnetic field direction in strongly magnetized molecular cloud environments (Fiedler & Mouschovias 1993; Galli & Shu 1993; Allen et al. 2003). On the smaller star formation scale of ∼1000 au, simulations show that both a magnetic field and rotation are essential for launching, collimating, and stabilizing the jet/outflow (Gardiner et al. 2000; O'Sullivan & Ray 2000; Pudritz & Banerjee 2005). Without the inclusion of a magnetic field in their simulations, no stabilized collimated jet/outflow can be made.

Statistical studies of the alignment between outflow and magnetic field directions on both large and small scales have been undertaken, but have not yet produced a conclusive result. For example, Davidson et al. (2011) find good alignment between the magnetic field and outflow directions in the young stellar objects of L1527 and IC348-SMM2. Hull et al. (2013) find no tendency for magnetic field and outflow directions to align, instead finding perpendicular or random orientation of magnetic fields with respect to outflow directions based on observations of 16 low-mass protostars (Chapman et al. 2013). Another recent study toward 62 low-mass young stellar objects has shown that magnetic field directions are misaligned with outflow directions by $50° \pm 15°$ in three-dimensional space, rather than being randomly oriented with respect to one other (Yen et al. 2021).

Magnetic fields have also been studied toward several outflows using molecular line polarization observations. Their results show that magnetic field directions are either parallel or perpendicular to outflow directions. For example, Glenn et al. (1997) found that the position angles of HCO$^+$ (1−0) line polarization vectors are $47° \pm 5°$, which is about 20° offset with respect to the DR21 outflow direction. Cortes et al. (2006) found that $^{12}$CO (2−1) line polarization vectors show position angles of $31°.5 \pm 8°.7$ and $48°.5 \pm 7°.9$ in the redshifted velocity components and $-50°.2 \pm 9°.2$ and $53°.6 \pm 8°.9$ in the blueshifted velocity components of the NGC 2071IR outflows. The derived polarization vectors are either parallel or perpendicular to the outflow direction of ∼40°–50°. Lee et al. (2018a) found that the position angles of SiO (8−7) line polarization vectors are mainly parallel to the HH 211 jet axis. Ching et al. (2016) found that $^{12}$CO (3−2) line polarization vectors have a 20° offset relative to the NGC1333 IRAS 4A outflow direction. However, there is an ambiguity in the determination of a magnetic field direction due to the Goldreich–Kylafis Effect (Goldreich & Kylafis 1981). Molecular line polarization can be either parallel or perpendicular to the magnetic field direction, depending on the angles between the magnetic field and the line of sight. Meanwhile, it is quite straightforward to study the magnetic field morphology using dust polarized emission at submillimeter wavelengths, since the magnetic field is perpendicular to the polarization vector.

One of the BISTRO targets, NGC 2071IR, has a massive core with several near-infrared objects and nebulae in its central region (Walther et al. 1993; Tamura et al. 2007; Walther & Gaballe 2019).

The distance to our target is $417 \pm 5$ pc, adopted from the measured distance to NGC 2068 using Gaia data (Kounkel et al. 2018). Within the region, there are many outflows including a powerful northeast–southwest bipolar outflow. IRS 3 ($\sim 1\,M_\odot$; Trinidad et al. 2009) has been identified as the launching source of this large-scale outflow, which extends over $\sim 0.6$ pc (Bally 1982; Snell et al. 1984; Eislöffel 2000).

A polarization study toward NGC 2071IR has previously been performed by Matthews et al. (2002) using the SCUPOL polarimeter for SCUBA on the JCMT. They found a pinched magnetic field morphology in the central region, which has also been seen in several other collapsing star-forming regions (e.g., Girart et al. 2006; Attard et al. 2009; Rao et al. 2009; Hull et al. 2014; Cox et al. 2018; Lee et al. 2018b; Maury et al. 2018; Sadavoy et al. 2018; Kwon et al. 2019). However, Matthews et al. (2002) note that the pinched magnetic field structure of the NGC 2071IR core might be different from other hourglass shapes which are produced via a dragged magnetic field by the infalling material, since the core does not show any internal flattened shape in their continuum map. They derive a magnetic field strength of 56 $\mu$G in the dense core region using the Davis–Chandrasekhar–Fermi method (Davis 1951; Chandrasekhar & Fermi 1953). Their study also shows that the polarization fraction decreases with intensity with a power-law slope of $-0.79$. A polarization study toward the IRS 3 bipolar outflow has been also undertaken by Cortes et al. (2006) using the Berkeley–Illinois–Maryland Association (BIMA) array with a spatial resolution of about $4''$. They found that magnetic field vectors inferred from $^{12}$CO (2−1) line polarization observations are parallel to outflow directions.

Thanks to the high sensitivity of POL-2, we can investigate the magnetic field structure in detail not only within the central region but also across the whole area of NGC 2071IR over which the bipolar outflow spreads. Simultaneously obtained 450 $\mu$m and 850 $\mu$m continuum polarization data also provide an opportunity to study dust grain properties.

The paper is organized as follows. In Section 2, we describe our observations and the reduction of the JCMT POL-2 polarization data. In Section 3, we present the results of our 450 $\mu$m and 850 $\mu$m continuum polarization observations. We derive the magnetic field strength from polarization angles measured at 450 $\mu$m and 850 $\mu$m using the modified Davis–Chandrasekhar–Fermi method, and study grain properties using 450 $\mu$m and 850 $\mu$m polarization fractions. We also investigate the pinched magnetic field morphology and the alignment of the magnetic field direction with respect to the outflow. We finally compare the magnetic energy with the turbulent, gravitational, and outflow energies. In Section 4, we summarize our results. In the Appendix, we describe how we derive IRS 3 bipolar outflow kinetic energy density using HARP $^{12}$CO (3−2) and $^{13}$CO (3−2) molecular line data.

## 2. Observations and Data Reduction

Simultaneous 450 $\mu$m and 850 $\mu$m dust polarization observations were performed toward NGC 2071IR, with a reference position of R. A. = $05^h47^m05^s.040$, decl. = $00°21'51''.7$ (J2000) using POL-2/SCUBA-2 on the JCMT. The NGC 2071IR region was observed twenty times between 2016 September 8 and 2017 November 11 in Band 2 weather ($0.05 < \tau_{225\,\mathrm{GHz}} < 0.08$) for a total on-source time of 13.3 hr, using the POL-2-DAISY observing mode which produces maps with a diameter of $12'$ (Friberg et al. 2016). The spatial resolutions are $14''.1$ (corresponding to $\sim 5800$ au at a distance of 417 pc) at 850 $\mu$m and $9''.6$ ($\sim 4000$ au) at 450 $\mu$m. The achieved rms noise level in the 850 $\mu$m Stokes $I$ map is $\sim 2.0$ mJy beam$^{-1}$ on $12''$ pixels within the central $3'$ diameter area of constant exposure time. The rms noise level in the 850 $\mu$m Stokes $Q$ and $U$ maps is $\sim 0.8$ mJy beam$^{-1}$. The achieved rms noise level in 450 $\mu$m Stokes $I$ is $\sim 6.5$ mJy beam$^{-1}$ on $12''$ pixels, again within the central $3'$ diameter area. The rms noise levels in 450 $\mu$m Stokes $Q$ and $U$ are $\sim 2.9$ mJy beam$^{-1}$. We note that the derived noises are associated with the instrument, observing technique (Bastien et al. 2005a, 2005b; Holland et al. 2013; Friberg et al. 2016) and data reduction process (Berry et al. 2005; Chapin et al. 2013).

The POL-2 data reduction process uses the *pol2map* script recently added to SMURF (Berry et al. 2005; Chapin et al. 2013). The raw timestream data are first converted into separate Stokes $I$, $Q$, and $U$ timestreams using the process *calcqu*. We then produce an initial coadded $I$ map from a set of Stokes $I$ maps created from the Stokes $I$ timestreams of each observation using the iterative mapmaker *makemap*.

Final $I$, $Q$, and $U$ maps are produced using the *skyloop* task, which is a script that runs *makemap* simultaneously on the full set of 20 observations in order to find a solution that minimizes residuals across the full set of maps. The initial Stokes $I$ map is used to define a "mask" identifying areas containing astrophysical signals; see Mairs et al. (2015) for a detailed description of the role of masking in SCUBA-2 data reduction. The final coadded set of maps is corrected to account for any small differences in pointing between the input observations. The final $I$ map is used to estimate the $Q$ and $U$ signal caused by instrumental polarization (IP) at each point on the sky, using the "2019 August" IP model provided by the observatory.[86] The IP is approximately 1.5% and 2.3% at 450 $\mu$m and 850 $\mu$m, respectively, with a small dependence on elevation. The IP has been subtracted from all observations.

We also remove $^{12}$CO (3−2) emission contamination from the final 850 $\mu$m $I$ map following the process described by Parsons et al. (2018), since NGC 2071IR has a strong CO molecular bipolar outflow covering the whole region (Drabek et al. 2012). For subtracting the CO emission, we use archival JCMT HARP (Heterodyne Array Receiver Program) $^{12}$CO (3−2) data (project code MJLSG11) observed on 2007 November 25. We also use $^{13}$CO (3−2) and C$^{18}$O (3−2) molecular emission lines from these archival data to analyze the IRS 3 bipolar outflow.

The final $I$, $Q$, and $U$ maps, with a pixel size of $4''$, are combined using the task *pol2stack* to produce an output polarization vector catalog. All maps are calibrated in units of Jy beam$^{-1}$, using flux conversion factors (FCFs) of 725 mJy pW$^{-1}$ at 850 $\mu$m and 962 mJy pW$^{-1}$ at 450 $\mu$m (Friberg et al. 2016). Both FCF values take into account the additional losses of a factor of 1.35 at 850 $\mu$m and 1.96 at 450 $\mu$m due to the insertion of the POL-2 polarimeter into the SCUBA-2 light path. For 450 $\mu$m data, we convolve the 450 $\mu$m maps at each $4''$ pixel with a $14''.1$ Gaussian beam using the SMOOTH450 parameter in the *pol2map* task. This convolution procedure converts the beam resolution of the original 450 $\mu$m map, $9''.6$, to a resolution of $14''.1$, which allows us to make a direct comparison between the 450 $\mu$m and 850 $\mu$m data. We bin every $3 \times 3$ pixels with the $4''$ size into a pixel with a $12''$ size to improve the signal-to-noise ratio. All further analyses in this paper have been undertaken using these final maps with a pixel size of $12''$.

---
[86] https://www.eaobservatory.org//jcmt/wp-content/uploads/sites/2/2019/08/pol2ipcor.pdf

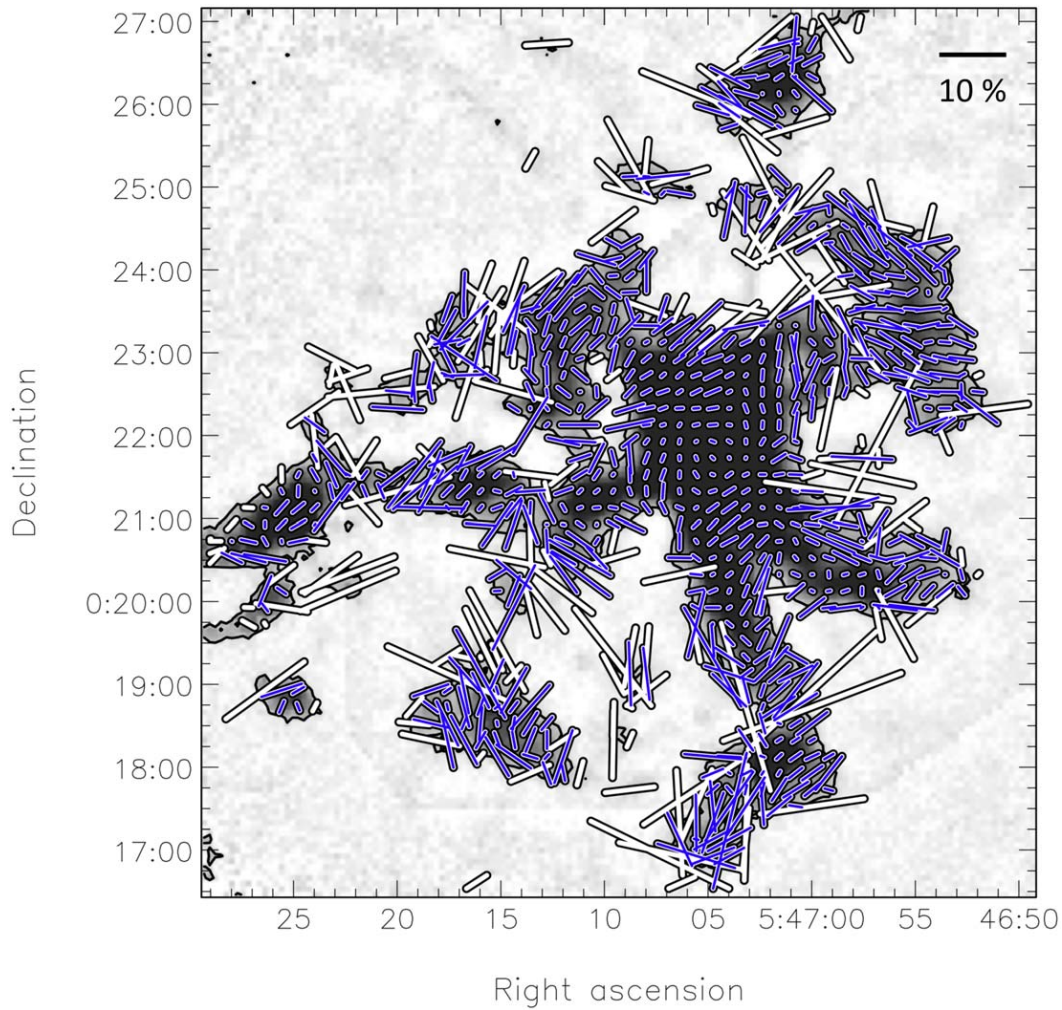

**Figure 1.** POL-2 850 $\mu$m polarization vector map of NGC 2071IR. Grayscale represents 850 $\mu$m continuum emission. The black contour corresponds to 15$\sigma$, where $\sigma$ = 2.0 mJy beam$^{-1}$. White vectors have a selection criterion of $(I/\delta I) \geqslant 10$, while blue vectors have a criterion of $(I/\delta I) \geqslant 20$. Vector lengths are proportional to their polarization fractions. A representative vector with a polarization fraction of 10% is shown in the upper right-hand corner.

The polarization angle $\theta$ and polarization fraction $P$ values used in this paper follow the conventional definitions of $\theta = 0.5 \arctan(U/Q)$ and $P = (Q^2 + U^2)^{0.5}/I$, respectively. Polarization angle is measured from the north, increasing eastward. The debiased polarization fraction is $P_{db} = (Q^2 + U^2 - 0.5[(\delta Q)^2 + (\delta U)^2])^{0.5}/I$. The debiased polarization fraction uncertainty is $\delta P_{db} = (P_{db}I^2)^{-1}\sqrt{Q^2\delta Q^2 + U^2\delta U^2 + P_{db}^4 I^2\delta I^2}$, where $\delta I$, $\delta Q$, and $\delta U$ are measurement errors in those maps. We use the debiased data for all magnetic field studies, while we use the nondebiased data when using a Rician noise model to study the relationship between polarization fraction and total intensity. More detail on the data reduction process is given by Pattle et al. (2021). Figure 1 shows the final POL-2 850 $\mu$m polarization vector map, with vector selection criteria of $(I/\delta I) \geqslant 10$ (white vectors) and $(I/\delta I) \geqslant 20$ (blue vectors). Figure 2 shows the magnetic field vector map, with each 850 $\mu$m polarization vector rotated by 90°, with vector selection criteria of $(I/\delta I) \geqslant 10$ and $(P/\delta P) \geqslant 3$. Figure 3 shows the 450 $\mu$m polarization vector map (left panel) and magnetic field vector map (right panel) with vector selection criteria of $(I/\delta I) \geqslant 50$ and $(P/\delta P) \geqslant 3$.

Figure 4 shows, for the central region, a comparison of polarization angles and fractions at 450 $\mu$m and 850 $\mu$m, with a pixel size of 12″. We note that the polarization angles at the two wavelengths show a good agreement in the central dense region.

We take 220 GHz continuum and C$^{18}$O (2−1) molecular line emission data from the Submillimeter Array (SMA) archive (Project code 2009B-S004) to study the kinematics in the central region. The SMA observation was carried out using the compact configuration, with projected baselines ranging from 8 to 76 m. The imaging process is undertaken using the Miriad software. The synthesized map using the natural weighting has a beam size of 3″.4 × 2″.8, with a position angle of −5°. Three continuum emission peaks are detected, with peak intensities of ∼0.22 Jy for IRS 1, ∼0.09 Jy for IRS 2, and ∼0.37 Jy for IRS 3.

## 3. Results

### 3.1. Magnetic Field Strength Using the DCF Method

Knowledge of magnetic field strength is crucial in order to determine whether magnetic fields are important in star formation processes. We derive the plane-of-sky magnetic field strength ($B_{pos}$) from the 450 μm and 850 μm polarization data using the modified Davis–Chandrasekhar–Fermi method (DCF) (Davis 1951; Chandrasekhar & Fermi 1953) provided by Crutcher et al. (2004). Their equation is

$$B_{pos} = Q\sqrt{4\pi\rho}\,\frac{\sigma_\nu}{\sigma_\theta} \approx 9.3 \frac{\sqrt{n(H_2)}\,\Delta\nu}{\sigma_\theta}\,\mu G \quad (1)$$

where $N(H_2)$ is the volume density of molecular hydrogen in cm$^{-3}$; $\rho = \mu m_H N(H_2)$ in g cm$^{-3}$, where $\mu = 2.86$ is the mean molecular weight per hydrogen molecule (Kirk et al. 2013); $\sigma_\nu$ is the velocity dispersion in km s$^{-1}$; $\Delta\nu = \sigma_\nu\sqrt{8\ln 2}$ is the FWHM velocity dispersion in km s$^{-1}$; $\sigma_\theta$ is the polarization angle dispersion in degrees; and $Q$ is a factor to account for the unresolved complex magnetic field and density structure within the beam size, which we take to be 0.5 as suggested by Ostriker et al. (2001) (see also Pattle et al. 2017).

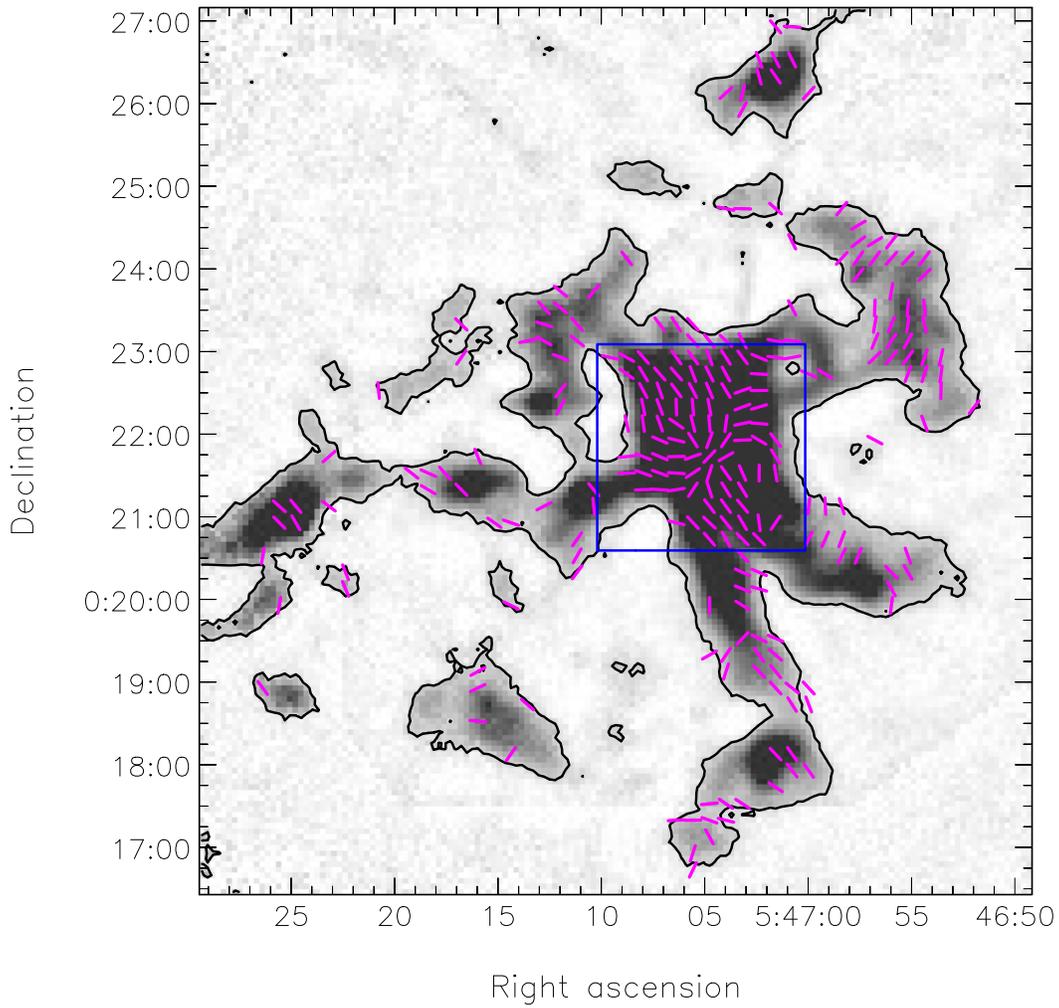

**Figure 2.** Magnetic field vector map, with each 850 μm polarization vector rotated by 90°, of NGC 2071IR. Vectors are selected using the criteria ($I/\delta I$) ⩾ 10 and ($P/\delta P$) ⩾ 3. Black contours correspond to 10σ, where σ = 2.6 mJy beam$^{-1}$. The blue box represents the central 2′.5 × 2′.5 region, within which an angle dispersion is derived for magnetic field strength measurement using the DCF method.

To determine the magnetic field strength, we first derive a polarization angle dispersion of 10°.7 ± 4°.9 within the central 2′.5 × 2′.5 region (marked as a blue box in Figure 2) using the unsharp masking method developed by Pattle et al. (2017). Following their procedure, we calculate a mean polarization angle in a 3 × 3 pixel box (corresponding to 36″ × 36″) using the relation $\bar{\theta} = 0.5\arctan(\bar{U}/\bar{Q})$, where the barred quantities, $\bar{U}$ and $\bar{Q}$, represent the average values of these quantities within the box. We then calculate an angle dispersion of $|\theta_i - \bar{\theta}_i|$ within the central 2′.5 × 2′.5 region using the equation

$$\sigma_\theta = \sqrt{\frac{1}{N}\sum_{i=1}^{N}(|\theta_i - \bar{\theta}_i| - \overline{|\theta_i - \bar{\theta}_i|})^2} \quad (2)$$

where $\theta_i$ is the polarization angle in the $i$th 12″ × 12″ pixel, and $\bar{\theta}_i$ is the mean polarization angle in the 3 × 3 pixel box centered on the $i$th pixel. The angle dispersion uncertainty of 4°.9 is from the median value of angle uncertainties of the selected polarization vectors.

The nonthermal velocity dispersion, $\sigma_\nu \sim 0.45$ km s$^{-1}$, is derived from the C$^{18}$O (3−2) molecular line velocity dispersion obtained with HARP in the central region, $\sigma_{\nu,C^{18}O} \sim 0.46$ km s$^{-1}$, using the equation

$$\sigma_\nu^2 = \sigma_{\nu,C^{18}O}^2 - \frac{kT}{m_{C^{18}O}} \quad (3)$$

where $k$ is the Boltzmann constant. The adopted temperature is 20.1 ± 1.8 K, estimated using Herschel data (Könyves et al. 2020). Therefore, the FWHM corresponding to the nonthermal velocity dispersion is $\Delta \nu = \sigma_\nu \sqrt{8 \ln 2} \sim 1.07$ km s$^{-1}$. This value is almost identical to the FWHM of NH$_3$(1,1) molecular line measured by Takano et al. (1986).

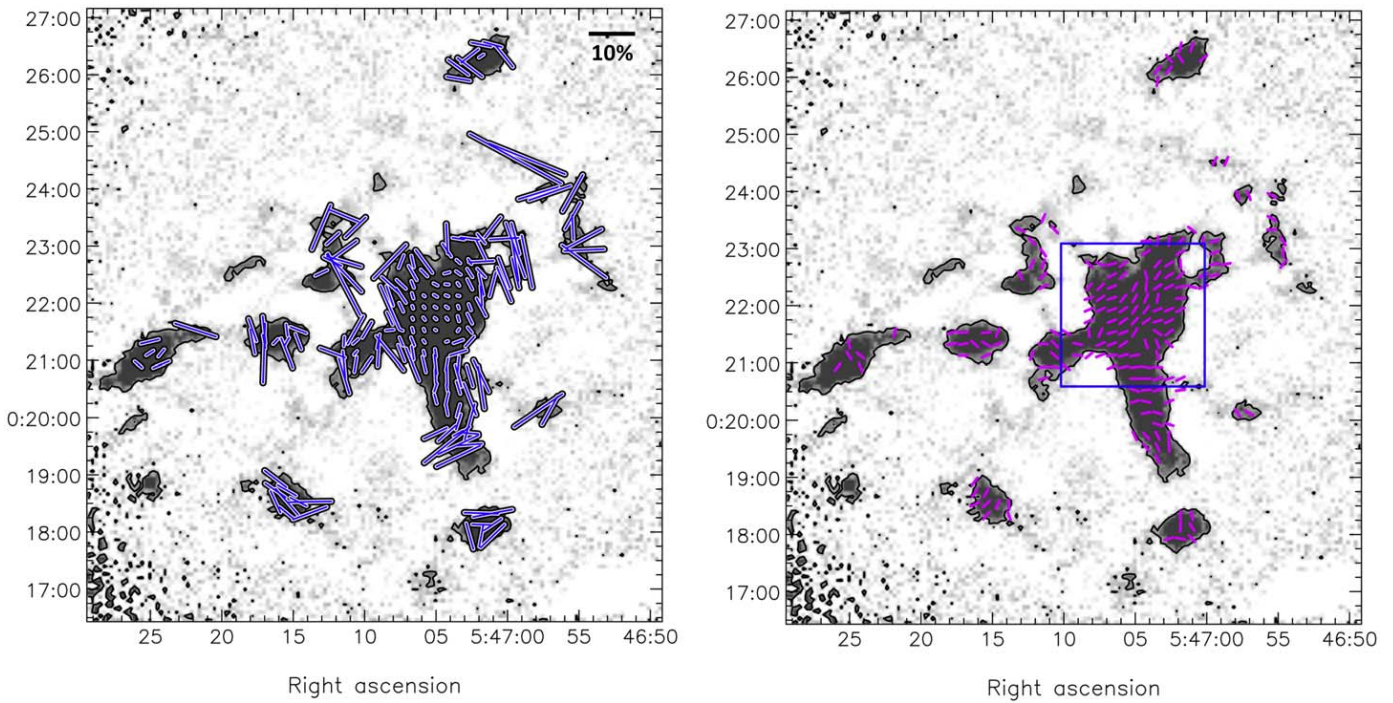

**Figure 3.** The left-hand panel shows the POL-2 450 μm polarization vector map. Vectors have a selection criterion of $(I/\delta I) \geqslant 50$ and $(P/\delta P) \geqslant 3$. Vector lengths are proportional to their polarization fractions. A representative vector with a polarization fraction of 10% is shown in the upper right-hand corner. The right-hand panel shows the magnetic field vector map, with each 450 μm polarization vector rotated by 90°. Grayscale represents 450 μm continuum emission. The black contour corresponds to 20σ, where $\sigma = 6.5$ mJy beam$^{-1}$. The blue box represents the central $2.'5 \times 2.'5$ region, within which an angle dispersion is derived for magnetic field strength measurement using the DCF method.

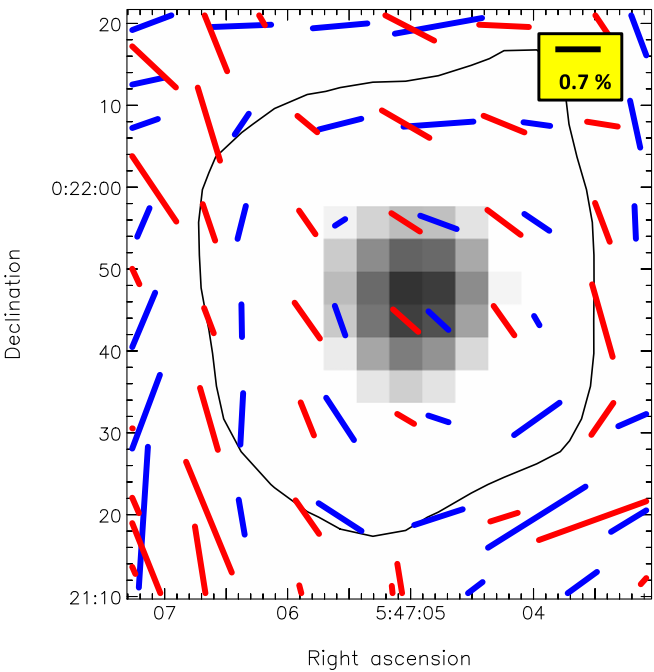

**Figure 4.** Comparison between 450 μm (red) and 850 μm (blue) polarization vectors in the central $60'' \times 70''$ area. All polarization vectors are selected using the criterion of $(P/\delta P) \geqslant 3$ at both wavelengths. A representative vector with a polarization fraction of 0.7% is shown in the upper right-hand corner. The black contour represents the 450σ level of 850 μm continuum emission, where $\sigma = 2.0$ mJy beam$^{-1}$.

We next derive magnetic field strengths toward the same central point using two differently sized regions. One is for the central $60'' \times 60''$ region, representing a linear size of 0.12 pc, and the other is for the central $100'' \times 100''$ region, with a linear size of 0.2 pc. In the first compact core region, we obtain a column density of $N(H_2) = 1.08(\pm 0.62) \times 10^{23}$ cm$^{-2}$ based on the measured total flux of $31.05 \pm 3.42$ Jy at 850 μm and adopting a dust opacity range of $\kappa_{850\,\mu m} = 0.018 \pm 0.009$ cm$^2$ g$^{-1}$, assuming a gas-to-dust ratio of 100 (Ossenkopf & Henning 1994). The derived average volume density and total mass in this region are $n(H_2) = 3.67(\pm 2.12) \times 10^5$ cm$^{-3}$ and $M = 36.3 \pm 21.0 \, M_\odot$, respectively, assuming a uniform-density cylindrical shape with a length of $L = 0.12$ pc. We then derive a plane-of-sky magnetic field strength of $563 \pm 421$ μG using Equation (1).

In the central $100'' \times 100''$ area, representing a volume 4.5 larger than that of the $60'' \times 60''$ area, we obtain a column density of $N(H_2) = 4.61(\pm 2.67) \times 10^{22}$ cm$^{-2}$ based on the measured total flux of $36.9 \pm 4.1$ Jy at 850 μm. The derived average volume density and total mass are $N(H_2) = 9.41(\pm 5.34) \times 10^4$ cm$^{-3}$ and $M = 43.1 \pm 24.9 \, M_\odot$, respectively, assuming a uniform-density cylindrical shape with a length of $L = 0.2$ pc. The estimated average plane-of-sky magnetic field strength is $285 \pm 212$ μG.

We adopt $B_{pos} = 563 \pm 421$ μG as a representative magnetic field strength in NGC 2071IR, since the central $60'' \times 60''$ region holds most of the mass of NGC 2071IR (see Table 1). The $100'' \times 100''$ region is only 20% more massive than the central $60'' \times 60''$ region, despite being 4.5 times larger in volume. We additionally derive an angular dispersion of $\sigma_\theta = 14.°0 \pm 5.°0$ in the 450 μm polarization data, again using the unsharp masking method, within the central $2.'5 \times 2.'5$ region. The plane-of-sky magnetic field strength is therefore calculated to be about $429 \pm 278$ μG using the 450 μm polarization data. Since the rms noises of our $Q$ and $U$ measurements at 850 $\mu$m are about three times smaller than those at 450 $\mu$m, we take $B_{pos} = 563 \pm 421$ $\mu$G measured at 850 $\mu$m as a representative magnetic field strength in NGC 2071IR.

**Table 1**
Parameters Used for the Measurement of Magnetic Field Strength Using DCF Method

| Parameter | Value |
| --- | --- |
| $\sigma_\theta$ (polarization angle dispersion) | $10°.7 \pm 4°.9$ |
| $\Delta\nu$ (FWHM nonthermal velocity dispersion) | $1.07$ km s$^{-1}$ |
| $n(H_2)$ (volume density of molecular hydrogen) | $3.67(\pm 2.12) \times 10^5$ cm$^{-3}$ |
| $N(H_2)$ (column density in the central 0.12 pc area) | $1.08(\pm 0.62) \times 10^{23}$ cm$^{-2}$ |
| $\rho$ (mass density with a molecular weight $\mu = 2.86$) | $1.75(\pm 1.01) \times 10^{-18}$ g cm$^{-3}$ |
| $B_{pos}$ (plane-of-sky magnetic field strength) | $563 \pm 421$ $\mu$G |

Our derived magnetic field strength, 56 $\mu$G, is about ten times stronger than the strength derived in Matthews et al. (2002). Their smaller magnetic field strength is because, compared with our values, they used a larger derived polarization angle dispersion of 33° and a smaller volume density of $N(H_2) \sim 10^4$ cm$^{-3}$, based on the critical density of the $NH_3(1,1)$ molecular line. Our adopted unsharp masking method for the polarization angle dispersion measurement (Pattle et al. 2017) provides a smaller value than the one measured with the method in Matthews et al. (2002). This is because the unsharp method enables us to measure the angle dispersion along a curved mean field direction, while the method used in Matthews et al. (2002) assumes a uniform mean field direction. Our derived volume density of $3.67(\pm 2.12) \times 10^5$ cm$^{-3}$ is also more likely to be accurate than the density value assumed from ammonia detection.

In order to compare our polarization data obtained using POL-2 with those of Matthews et al. (2002) using SCUPOL, we measure polarization angle dispersion and mean polarization angle from POL-2 data using the same method used by them. We derive a polarization angle dispersion of $\sigma_\theta = 38°$ and a mean polarization angle of $\bar\theta = 33°$, respectively. These obtained values are almost the same as theirs, which are $\sigma_\theta = 33°$ and $\bar\theta = 34°$, respectively. This result confirms that both data sets are consistent with each other at the central strong emission region.

### 3.2. Polarization Properties

We investigate the debiased and nondebiased polarization fractions as functions of total intensities at both 850 $\mu$m and 450 $\mu$m using two different fitting methods, the single power law, and the Rician-mean model. We used the following power-law model

$$P = P_{\sigma_{QU}}(I/\sigma_{QU})^{-\alpha} \quad (4)$$

where $\sigma_{QU}$ is the rms noise in both Stokes $Q$ and $U$ and $P_{\sigma_{QU}}$ is the polarization fraction at the noise level of the data. Since the rms noises of the $Q$ and $U$ measurements are almost the same, a single mean value $\sigma_{QU}$ representing both noises is chosen. The polarized intensity $I_p = (Q^2 + U^2)^{0.5}$ mathematically follows a Rician distribution once both Stokes $Q$ and $U$ have Gaussian distributions. We used Equation (21) of Pattle et al. (2019) as the Rician-mean model. Pattle et al. (2019) showed that the Rician-mean model can accurately recover the power law up to an index of ~0.6, while the single power-law model can only recover indices less than 0.3. We define the goodness-of-fit parameter

$$\chi^2 = \sum_{i=1}^{N} \left(\frac{P_i^m - P_i^{obs}}{\delta P_i^{obs}}\right)^2, \quad (5)$$

where $P_i^{obs}$ and $\delta P_i^{obs}$ are the observed polarization fraction and its uncertainty, respectively, $P_i^m$ is the polarization fraction calculated from the power-law model or the Rician-mean model, and $N$ is the number of data. By minimizing $\chi^2$, we find the best-fitting parameters $\alpha$ and $P_{\sigma_{QU}}$ for the power-law and Rician-mean models. Table 2 lists the fitting results.

Figure 5 shows the debiased and nondebiased polarization fractions as a function of total intensity at 850 $\mu$m within the central 3' area. In the left panel of Figure 5, we fit the debiased polarization fraction as a function of the normalized total intensity with the power law. We obtain best-fit parameters of $\alpha = 0.35 \pm 0.03$ and $P_{\sigma_{QU}} = 0.09 \pm 0.02$ using 158 polarization vectors selected using the criterion of $(I/\delta I) \geqslant 10$, where $\sigma_{QU} = 0.80$ mJy beam$^{-1}$. For 96 polarization vectors selected using the criterion of $(P/\delta P) \geqslant 3$, the best-fit parameters for the power-law relation are $\alpha = 0.43 \pm 0.04$ and $P_{\sigma_{QU}} = 0.17 \pm 0.05$. The right-hand panel of Figure 5 shows the Rician-mean model fitting to the nondebiased polarization fraction as a function of total intensity for 166 data points without any criterion condition. We obtain the best Rician-mean model fitting parameters $\alpha = 0.36 \pm 0.04$ and $P_{\sigma_{QU}} = 0.10 \pm 0.03$. The power-law index, $0.35 \pm 0.03$, fitted to debiased data points with the selection criterion based on the total intensity ($(I/\delta I) \geqslant 10$), is almost the same as the slope value, $0.36 \pm 0.04$, obtained from the Rician-mean model fitting to the whole nondebiased data set (see Table 2). However, the slope of the power law, $0.43 \pm 0.04$, fitted to the data points (filled circles in the left panel of Figure 5) selected with the $(P/\delta P) \geqslant 3$ criterion, is steeper than the slope from the Rician-mean model. This is because the selection criterion based on the polarization fraction introduces a systematic bias by excluding data points with lower polarization fractions as shown in the left-hand panel (see also Figure 2 in Pattle et al. 2019).

Figure 6 shows the debiased and nondebiased polarization fractions as a function of total intensity at 450 $\mu$m within the central 3' area. The left-hand panel of Figure 6 shows a single power-law fit to the debiased polarization fraction as a function of the normalized total intensity with $\sigma_{QU} = 2.93$ mJy beam$^{-1}$. We obtain best-fit parameters of $\alpha = 0.57 \pm 0.03$ and $P_{\sigma_{QU}} = 0.49 \pm 0.11$ for the power-law relation using 150 polarization vectors selected using the criterion of $(I/\delta I) \geqslant 10$. For 105 polarization vectors selected using the criterion of $(P/\delta P) \geqslant 3$, the best-fit parameters for the power-law relation are $\alpha = 0.63 \pm 0.03$ and $P_{\sigma_{QU}} = 0.82 \pm 0.17$. The right-hand panel of Figure 6 shows the Rician-mean model fitting to the full set of 164 data points without any selection criterion. The obtained best Rician-mean model fitting parameters are $\alpha = 0.59 \pm 0.03$ and $P_{\sigma_{QU}} = 0.55 \pm 0.12$ (see Table 2). The power-law index, $0.57 \pm 0.03$, fitted to debiased data points with the selection criterion based on the total intensity, is almost the same as the slope value, $0.59 \pm 0.03$, obtained from the Rician-mean model fitting to the whole nondebiased data set. As Pattle et al. (2019) pointed out, the Rician-mean model gives a best estimation of the $\alpha$ parameter from data without any selection criteria on non-debiased polarization fractions. We demonstrate that, from the comparisons of the slopes from the power-law and Rician-mean models, the slopes of the power-law model estimated from selected data, based on the polarization fraction criterion, have steeper slopes than those of the Rician-mean model.

**Table 2**
Fitting Results for the Rician-mean and Power-law Models

| Wavelength | $\sigma_{QU}$ | Rician-mean Model | | | | Power-law Model | | | | | | | |
|---|---|---|---|---|---|---|---|---|---|---|---|---|---|
| | | | | | | $(I/\delta I) \geqslant 10$ | | | | $(P/\delta P) \geqslant 3$ | | | |
| | | $\alpha$ | $P_{\sigma_{QU}}$ | $N$ | $\frac{\chi^2}{N-2}$ | $\alpha$ | $P_{\sigma_{QU}}$ | $N$ | $\frac{\chi^2}{N-2}$ | $\alpha$ | $P_{\sigma_{QU}}$ | $N$ | $\frac{\chi^2}{N-2}$ |
| 850 $\mu$m | 0.80 | 0.36 ± 0.04 | 0.10 ± 0.03 | 166 | 8.1 | 0.35 ± 0.03 | 0.09 ± 0.02 | 158 | 8.7 | 0.43 ± 0.04 | 0.17 ± 0.05 | 96 | 11.6 |
| 450 $\mu$m | 2.93 | 0.59 ± 0.03 | 0.55 ± 0.12 | 164 | 9.4 | 0.57 ± 0.03 | 0.49 ± 0.11 | 150 | 10.3 | 0.63 ± 0.03 | 0.82 ± 0.17 | 105 | 9.5 |

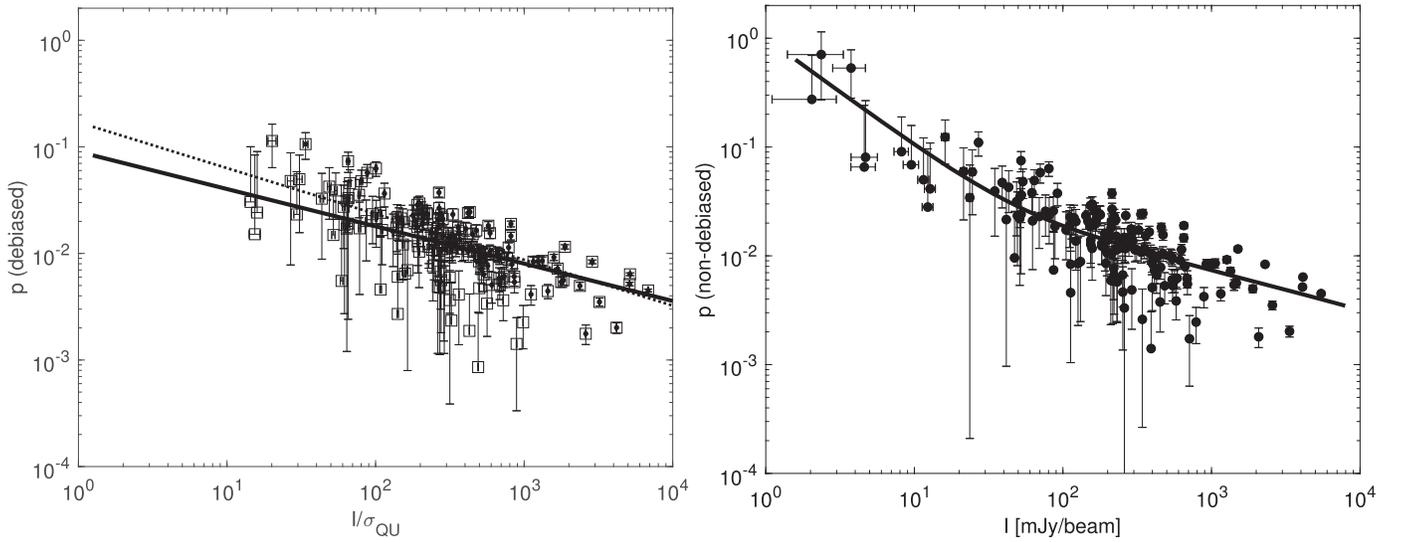

**Figure 5.** Polarization fraction as a function of total intensity at 850 $\mu$m on 12″ pixels within the central 3′ area. The left-hand panel shows a single power-law fit to the debiased polarization fraction as a function of the normalized total intensity with $\sigma_{QU} = 0.80$ mJy beam$^{-1}$. All squares represent a total of 158 vectors selected using the criterion $(I/\delta I) \geqslant 10$. The obtained power-law slope, $\alpha$, and $P_{\sigma QU}$ are $0.35 \pm 0.03$ and $0.09 \pm 0.02$ (solid line), respectively. Filled circles represent a total of 96 vectors, selected using the criterion $(P/\delta P) \geqslant 3$. The obtained power-law slope, $\alpha$, and $P_{\sigma QU}$ are $0.43 \pm 0.04$ and $0.17 \pm 0.05$ (dotted line), respectively. The right-hand panel shows the Rician fitting to the nondebiased polarization fraction as a function of the total intensity for the whole data set within the central 3′ region. The derived $\alpha$ and $P_{\sigma QU}$ are $0.36 \pm 0.04$ and $0.10 \pm 0.03$, respectively (see details about Rician fitting in Pattle et al. 2019). We note that negative error bars for some data points are not shown due to the logarithmic scale of the vertical axis. Some error bars are too small to be seen compared to the symbol size.

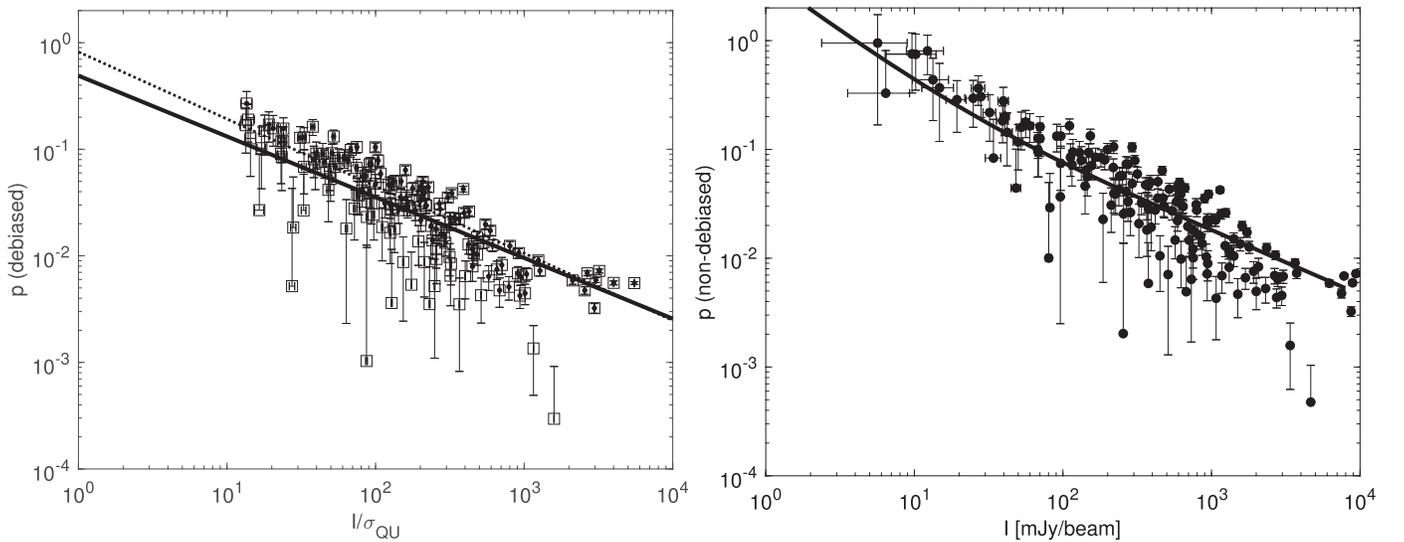

**Figure 6.** Polarization fraction as a function of total intensity at 450 $\mu$m on 12″ pixels within the central 3′ area. The left-hand panel shows a single power-law fit to the debiased polarization fraction as a function of the normalized total intensity with $\sigma_{QU} = 2.93$ mJy beam$^{-1}$. All squares represent a total of 150 vectors selected using the criterion $(I/\delta I) \geqslant 10$. The obtained power-law slope, $\alpha$, and $P_{\sigma QU}$ are $0.57 \pm 0.03$ and $0.49 \pm 0.11$ (solid line), respectively. Filled circles represent a total of 105 vectors, selected using the criterion $(P/\delta P) \geqslant 3$. The obtained power-law slope, $\alpha$, and $P_{\sigma QU}$ are $0.63 \pm 0.03$ and $0.82 \pm 0.17$ (dotted line), respectively. The right-hand panel shows the Rician fitting to the nondebiased polarization fraction as a function of the total intensity for the whole data set within the central 3′ region. The derived $\alpha$ and $P_{\sigma QU}$ are $0.59 \pm 0.03$ and $0.55 \pm 0.12$, respectively. We note that negative error bars for some data points are not shown due to the logarithmic scale of the vertical axis. Some error bars are too small to be seen compared to the symbol size.

We find that the polarization fraction decreases with intensity at both 450 $\mu$m and 850 $\mu$m. This trend of decreasing polarization fraction has often been shown in the dense regions of low-mass/massive star-forming molecular cores, such as Bok globules CB 54 and DC 253-1.6 (Henning et al. 2001), dense cores within the dark cloud Barnard 1 in Perseus molecular cloud (Matthews & Wilson 2002), massive molecular cores of W51 e1 and e2 (Lai et al. 2001), and giant molecular cloud complex Vela C (Fissel et al. 2016). The suggested possible explanations for this trend are due to: (1) the loss of the grain alignment because of the grain growth (Bethell et al. 2007; Brauer et al. 2016) or scattering and absorption of photons at high density (King et al. 2018, 2019); (2) the optical depth effect (Dowell 1997; Hildebrand et al. 1999); or (3) the unresolved complex $B$-field structures along the line of sight. It is, however, beyond the scope of this paper to constrain them using only these data.

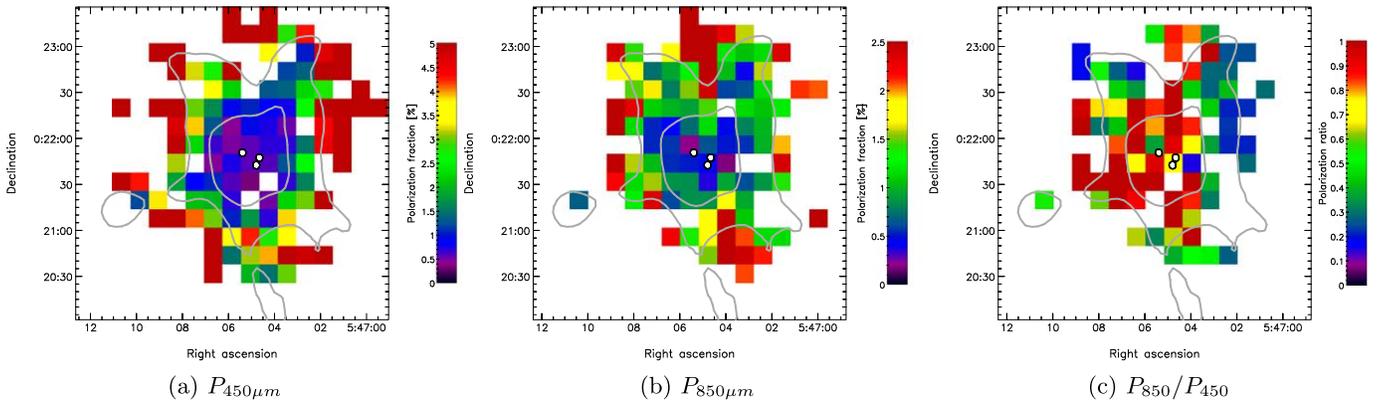

**Figure 7.** (a) Polarization fraction map at 450 $\mu$m. (b) Polarization fraction map at 850 $\mu$m. (c) Distribution of polarization fraction ratios of 850 to 450 $\mu$m, $P_{850}/P_{450}$. The gray contours represent the 130 $\sigma$ and 400$\sigma$ levels of 850 $\mu$m continuum emission, where $\sigma = 2.0$ mJy beam$^{-1}$. The three 220 GHz SMA continuum emission peaks, IRS 1, 2, and 3, are marked as small circles in all three panels.

We note that the Rician-mean model fitting slope of $0.36 \pm 0.04$ at 850 $\mu$m is much shallower than the slope of $0.59 \pm 0.03$ at 450 $\mu$m. We test whether it is due to the optical depth effect at 450 $\mu$m and 850 $\mu$m. The optical depth at 850 $\mu$m is smaller than that at 450 $\mu$m by a factor of $\tau_{850}/\tau_{450} = (\lambda_{450}/\lambda_{850})^\beta \sim 0.4$, where the adopted mean value of the dust opacity spectral index is $\beta \sim 1.4$ in the central $60'' \times 60''$ region. The $\beta$ value is estimated using the following equation, where the temperature is $20.1 \pm 1.8$ K and measured total fluxes are 168.63 Jy at 450 $\mu$m and 31.05 Jy at 850 $\mu$m, respectively. The derived optical depths are $\tau_{850} \sim 0.012$ and $\tau_{450} \sim 0.026$, respectively, both of which are optically thin. Therefore, the optical depth is not the main reason for the different slopes at 450 $\mu$m and 850 $\mu$m

$$\frac{I_{\nu_{450\,\mu m}}}{I_{\nu_{850\,\mu m}}} = \frac{e^{\frac{h\nu_{850\,\mu m}}{kT}}-1}{e^{\frac{h\nu_{450\,\mu m}}{kT}}-1}\left(\frac{\nu_{450\,\mu m}}{\nu_{850\,\mu m}}\right)^{\beta+3}. \quad (6)$$

The Rician-mean model fitting slope of $0.36 \pm 0.04$ obtained from the 850 $\mu$m polarization data is almost the same as the slope of 0.33 toward Oph A derived by Pattle et al. (2019). They suggest that the shallower slope of 0.33 in Oph A, compared to a slope of 0.76 toward both Oph B and C, could be due to strong radiation from the B spectral type star S1, which may help maintain grain alignment in the higher optical depth region of Oph A. As is the case in Oph A, NGC 2071IR shows better grain alignment than do Oph B and C. The central three infrared young stellar objects, IRS 1, 2, and 3 in NGC 2071IR could be radiation sources, which are assisting grain alignment.

We find that the median value of polarization fractions is 3.0% at 450 $\mu$m with the condition of $(P/\delta P) \geqslant 3$ in the central $3'$ region, which is larger than the median value of 1.2% at 850 $\mu$m. The ratio of the median polarization fractions at 850 and 450 $\mu$m, $P_{850}/P_{450}$, is 0.37. The trend of a smaller polarization fraction at a larger wavelength has been also observed in M17, shown by Zeng et al. (2013). They interpreted that it is due to the better alignment of warmer dust in the strong radiation environment. In Figure 7, we visualize the polarization fraction distributions at 450 $\mu$m and 850 $\mu$m and the polarization ratio of $P_{850}/P_{450}$. It is clearly shown that the polarization fractions at both wavelengths decrease as the intensity increases. Combining the result of a shallow Rician-mean model fitting slope of 0.36 at 850 $\mu$m, and the higher polarization fraction at 450 $\mu$m than 850 $\mu$m ($P_{850} < P_{450}$), we suggest that the central strong radiation from YSOs are indeed assisting the grain alignment suggested by the radiative alignment torques (RAT) theory (Lazarian & Hoang 2007).

Figure 7(c) shows the distribution of $P_{850}/P_{450}$ values, which are <1, slightly increasing toward the center. It is because the dust alignment persists more efficiently at 850 $\mu$m compared to 450 $\mu$m, as presented by their Rician-mean model fitting slopes of 0.36 at 850 $\mu$m and 0.59 at 450 $\mu$m. We note that there might be some changes in dust grain properties (e.g., grain growth) in the center where the $P_{850}/P_{450}$ ratio shows a small increase.

### 3.3. Pinched Magnetic Field Morphology

Figures 1 and 2 show the 850 $\mu$m polarization vector map and magnetic field vector map (rotated by 90° from the polarization vector map) of NGC 2071IR, respectively. We confirm the pinched magnetic field structure along a northeast and southwest direction around the center, as shown in the blue box of Figure 2. This pinched structure was also seen by Matthews et al. (2002). Pinched magnetic field structures have typically been interpreted to have been formed by the ambipolar diffusion process in dense star-forming cores (e.g., Girart et al. 2006; Attard et al. 2009; Rao et al. 2009; Hull et al. 2014; Cox et al. 2018; Lee et al. 2018b; Maury et al. 2018; Sadavoy et al. 2018; Kwon et al. 2019). Therefore, one interpretation of this pinched magnetic field morphology could be magnetically regulated core collapse. However, observational evidence of infalling material in this system is required to confirm this hypothesis.

Another possible interpretation is that the magnetic fields are shaped by a rotating disklike structure and bipolar outflows from the IRS 3 young stellar object, which could create an apparent hourglass morphology (see, e.g., the schematic diagram of L1448 IRS 2 shown in Figure 3 of Kwon et al. (2019)). In Figure 8(a), we show the JCMT HARP $^{12}$CO (3–2) and C$^{18}$O (3−2) molecular line emission maps. We find that C$^{18}$O (3−2) emission is distributed perpendicular to the direction of the IRS 3 bipolar outflow traced by the $^{12}$CO (3−2) line emission (black contours). In addition, we find a slight velocity gradient from northwest to southeast.

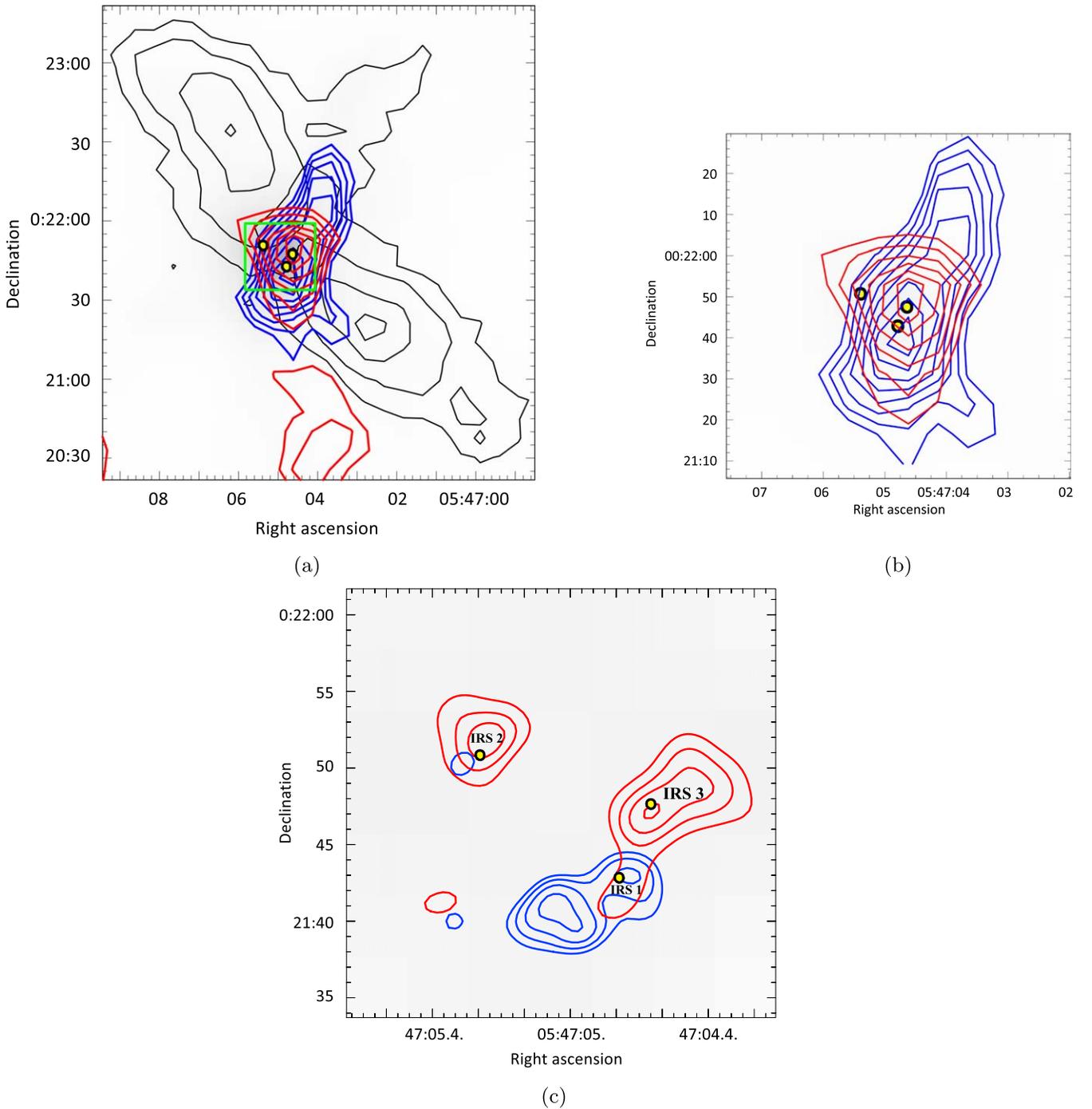

**Figure 8.** (a) JCMT HARP $^{12}$CO (3−2) and C$^{18}$O (3−2) molecular line emission. Blue and red contours represent the integrated C$^{18}$O (3−2) emission over the blueshifted velocity range of 7.5 ∼ 9.4 km s$^{-1}$ and the redshifted velocity range of 9.4 ∼ 11.0 km s$^{-1}$, respectively. The blue contour levels are 2.5, 2.7, 2.9, 3.1, 3.3, 3.5, 3.7, and 3.9 K km s$^{-1}$, and the red contour levels are 2.5, 2.9, 3.3, 3.7, 4.1, and 4.5 K km s$^{-1}$. Black contours represent the integrated $^{12}$CO (3−2) emission, which traces a bipolar outflow from IRS 3. The northeastern part is the integrated emission over the blueshifted velocity range of −45.3 ∼ −19.9 km s$^{-1}$, and the southwestern part is the integrated emission over the redshifted velocity range of 37.6 ∼ 63.0 km s$^{-1}$. Black contour levels are 3$\sigma$, 10$\sigma$, 30$\sigma$, and 50$\sigma$, where $\sigma$ = 0.7 K km s$^{-1}$. (b) The enlarged central region of the HARP C$^{18}$O (3−2) molecular line emission map. The peak position of the redshifted velocity component is northwest of the peak position of the blueshifted velocity component. (c) The SMA C$^{18}$O (2−1) molecular line emission in the central region, which is marked as a green box in (a). Blue and red contours represent the integrated C$^{18}$O (2−1) emission over the blueshifted velocity range of 4.0 ∼ 8.0 km s$^{-1}$ and the redshifted velocity range of 10.0 ∼ 14.0 km s$^{-1}$, respectively. The blue contour levels are 3.6$\sigma$, 4.4$\sigma$, 5.2$\sigma$, 6.0$\sigma$, and 6.8$\sigma$, and red contour levels are 3.6$\sigma$, 7.2$\sigma$, 10.8$\sigma$, and 14.4$\sigma$, where $\sigma$ = 0.25 Jy beam$^{-1}$ km s$^{-1}$. The three 220 GHz SMA continuum emission peaks, IRS 1, 2, and 3, are marked as small circles in all three figures.

Careful examination of the central part of Figure 8(b) shows that the peak emission of the redshifted C$^{18}$O (3−2) velocity component (red contours) is located northwest of the peak emission position of the blueshifted velocity component (blue contours). In Figure 8(c), the SMA C$^{18}$O (2−1) molecular line emission map of the very central region shows a clear velocity gradient from the northwest redshifted velocity component to the southeast blueshifted velocity component. This gives us a hint that there is a rotating disklike structure perpendicular to the bipolar outflow direction. In Figure 9, we present the magnetic field vector map, overlaying a cartoon showing our suggestion of the rotating disklike structure and bipolar outflow. The shaded blue and red colors schematically represent outflow cavity walls and the shaded gray color represents a disklike structure. We suggest that the pinched magnetic field morphology is formed by the outflow cavity walls together with the rotating disklike structure at the center.

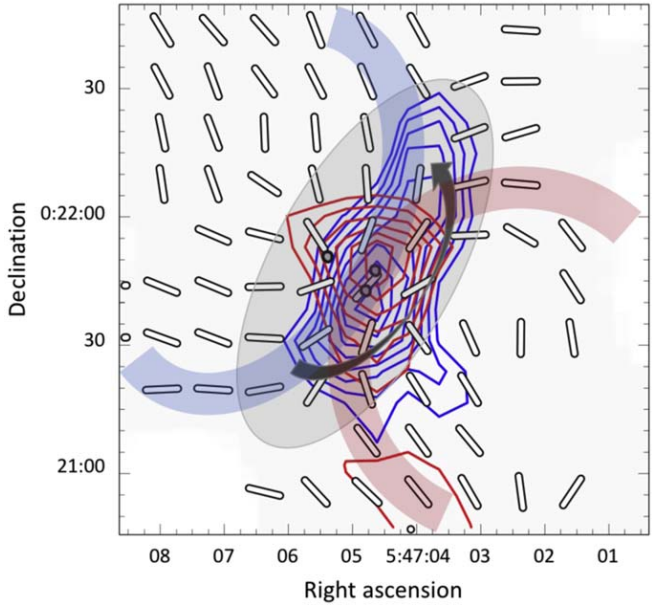

**Figure 9.** Magnetic field vectors derived from 850 μm polarization data using selection criteria of $(I/\delta I) \geq 10$ and $(P/\delta P) \geq 3$. The blue and red contours, which are the same as those in Figure 8, represent the HARP $C^{18}O$ (3−2) emission. We overlay a cartoon showing our suggestion of a rotating disklike structure and outflow. The shaded blue and red colors schematically represent outflow cavity walls, and the shaded gray color represents a disklike structure.

### 3.4. Alignment Between Outflow and Magnetic Field Directions

As discussed in Section 2, there is no conclusive consensus regarding the alignment between outflow and magnetic field directions. We consider NGC 2071IR to be an excellent environment in which to study the alignment issue. Thanks to the high sensitivity of our POL-2 map compared to the previous SCUPOL map presented by Matthews et al. (2002), we can investigate the magnetic field morphology in the outflow of IRS 3 in detail.

In Figure 10, we show the magnetic field vectors along with HARP $^{12}CO$ (3−2) and $^{13}CO$ (3−2) molecular line emission, which trace the bipolar outflow of IRS 3 well. In both panels, all of the magnetic field vectors shown are selected using the criteria of $(I/\delta I) \geq 10$ and $(P/\delta P) \geq 5$, derived from the 850 μm dust polarization data. In the left-hand panel, we present the high-velocity components of the $^{12}CO$ (3−2) molecular line emission, showing the blueshifted velocity range of $-45.3 \sim -19.9$ km s$^{-1}$ (blue contours) and the redshifted velocity range of $37.6 \sim 63.0$ km s$^{-1}$ (red contours). We do not use the low-velocity component of $^{12}CO$ (3−2) to present the outflow, since the low-velocity $^{12}CO$ (3−2) emission is contaminated by the ambient molecular cloud. In the right-hand panel, we present the low-velocity components of the $^{13}CO$ (3−2) molecular line emission, showing the blueshifted velocity range of $-1.1 \sim 5.1$ km s$^{-1}$ (blue contours) and the redshifted velocity range of $13.4 \sim 18.2$ km s$^{-1}$ (red contours).

In Figure 10, we show that the overall magnetic field direction is well aligned with the direction of the bipolar outflow. We also see the small-scale directional variation of magnetic field vectors over the region, like the curving shape in the blueshifted outflow component seen in both the $^{12}CO$ (3−2) and the $^{13}CO$ (3−2) emission maps. The directions of the magnetic field vectors are well aligned with the cavity walls shown in the low-velocity $^{13}CO$ (3−2) emission map in the right-hand panel of Figure 10. Due to this, we suggest that the magnetic field morphology of NGC 2071IR is formed by the strong influence of the outflow.

### 3.5. Comparison of Energy Densities

We compare the magnetic energy density with the turbulent, gravitational, and outflow energy densities of the NGC 2071IR star-forming region. First, we calculate the magnetic energy density using the equation

$$u_B = \frac{B^2}{8\pi}, \quad (7)$$

where the three-dimensional magnetic field strength, $B$, is estimated from the plane-of-sky magnetic field strength as $B \approx \frac{4}{\pi} B_{pos}$ assuming that, statistically, the three-dimensional magnetic field direction is isotropically distributed (Crutcher et al. 2004). The derived magnetic energy density is $u_B = 2.04(\pm 3.05) \times 10^{-8}$ erg cm$^{-3}$ using $B_{pos} = 563 \pm 421$ μG. Second, we find the turbulent kinetic energy density, $u_{turb} = \frac{3}{2}\rho v_{turb}^2 = 3.01(\pm 1.73) \times 10^{-8}$ erg cm$^{-3}$, adopting $v_{turb} = 1.07$ km s$^{-1}$ and $\rho = \mu m_H N(H_2) = 1.75(\pm 1.01) \times 10^{-18}$ g cm$^{-3}$. We adopt a volume density of $N(H_2) = 3.67(\pm 2.12) \times 10^5$ cm$^{-3}$, which is a value derived in the central $60'' \times 60''$ area. Third, we derive the gravitational energy density assuming a uniform-density sphere with a radius of $30''$. The estimated gravitational energy density is $u_g = 4\pi G \rho^2 R^2/5 = 1.81(\pm 1.51) \times 10^{-8}$ erg cm$^{-3}$, where $G$ is the gravitational constant. Finally, we obtain the kinetic energy density of the outflow, $u_{outflow} = 2.33(\pm 0.32) \times 10^{-8}$ erg cm$^{-3}$, using HARP $^{12}CO$ (3−2) and $^{13}CO$ (3−2) emission lines. We describe how we obtain the energy of the bipolar outflow using CO emission lines in the Appendix.

We find that the magnetic, turbulent kinetic, gravitational, and outflow kinetic energy densities are all comparable to each other in the NGC 2071IR star-forming region (see Table 3). We note, however, that the kinetic energy density of the outflow is a lower limit since we count only the radial velocity of the outflow. We do not have any information about the inclination of the outflow. Therefore, NGC 2071IR is in a state of energy equipartition now, even though the gravitational energy might have been the largest term at an earlier epoch in order to have formed stars in the central region.

### 4. Summary

We summarize the results of our 450 μm and 850 μm polarization observations toward the massive star-forming region NGC 2071IR as follows:

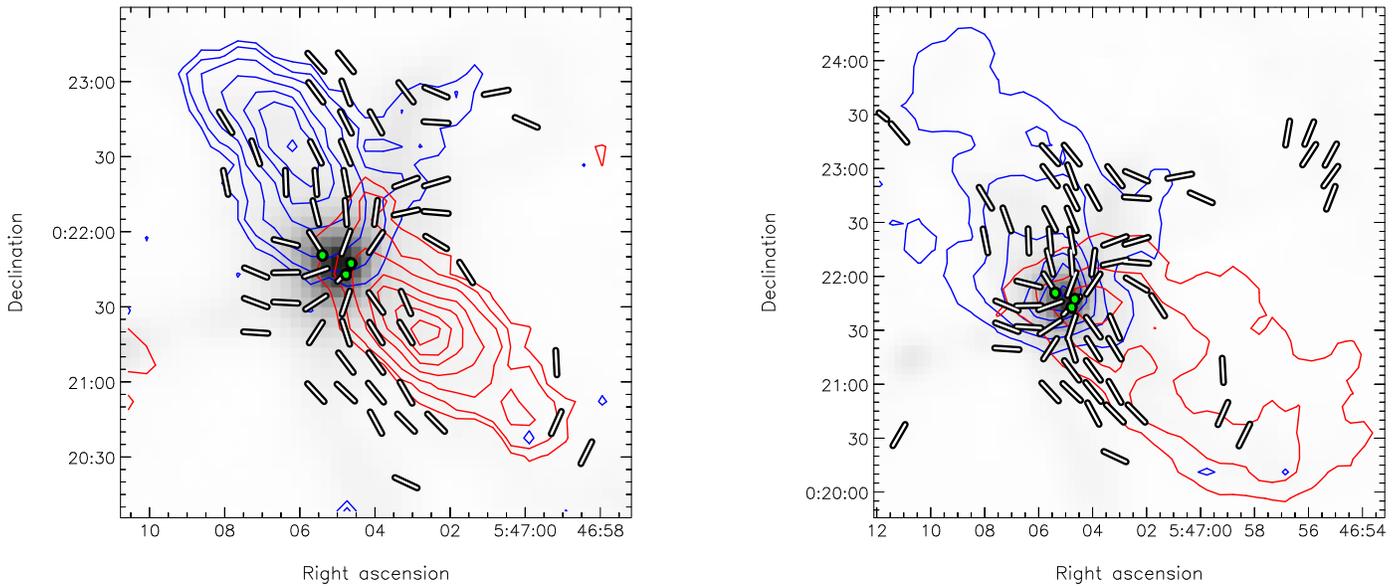

**Figure 10.** Magnetic field vector map with selection criteria of $(I/\delta I) \geqslant 10$ and $(P/\delta P) \geqslant 5$, derived from 850 $\mu$m dust polarization data. Left-hand panel: High-velocity components of the $^{12}$CO (3−2) molecular line emission. Blue and red contours represent the integrated $^{12}$CO (3−2) emission over the blueshifted velocity range of $-45.3 \sim -19.9$ km s$^{-1}$ and the redshifted velocity range of $37.6 \sim 63.0$ km s$^{-1}$, respectively. Contour levels are 3, 5, 10, 20, 30, 40, and 50$\sigma$, where $\sigma = 0.7$ K km s$^{-1}$. Right-hand panel: Low-velocity components of the $^{13}$CO (3−2) molecular line emission. Blue and red contours represent the integrated $^{13}$CO (3−2) emission over the blueshifted velocity range of $-1.1 \sim 5.1$ km s$^{-1}$ and the redshifted velocity range of $13.4 \sim 18.2$ km s$^{-1}$, respectively. Contour levels are 5, 15, 30, 50, 70, and 90$\sigma$, where $\sigma = 0.15$ K km s$^{-1}$. The three green points in the central area mark the positions of the IRS 1, 2, and 3 young stellar objects.

**Table 3**
Magnetic, Turbulent, Gravitational, and Outflow Energy Densities in Units of [erg cm$^{-3}$]

| $u_B$ | $u_{\text{turb}}$ | $u_g$ | $u_{\text{outflow}}$ |
|---|---|---|---|
| $2.04(\pm 3.05) \times 10^{-8}$ | $3.01(\pm 1.73) \times 10^{-8}$ | $1.81(\pm 1.51) \times 10^{-8}$ | $2.33(\pm 0.32) \times 10^{-8}$ |

1. We derive a plane-of-sky magnetic field strength of $B_{\text{pos}} = 563 \pm 421$ $\mu$G in the central $60'' \times 60''$ region from polarization angle dispersion of $10^\circ\!.7 \pm 4^\circ\!.9$ measured at 850 $\mu$m polarization data using the modified Davis–Chandrasekhar–Fermi method (Davis 1951; Chandrasekhar & Fermi 1953) provided by Crutcher et al. (2004). We also derive $B_{\text{pos}} = 429 \pm 278$ $\mu$G with the measurement of polarization angle dispersion, $14^\circ\!.0 \pm 5^\circ\!.0$, at 450 $\mu$m polarization data.

2. We find that the median value of polarization fractions is 3.0% at 450 $\mu$m in the central $3'$ region, which is larger than the median value of 1.2% at 850 $\mu$m. The trend could be due to the better alignment of warmer dust in the strong radiation environment (Zeng et al. 2013). We also find that polarization fractions decrease with intensity at both wavelengths, with slopes, determined by fitting a Rician noise model of $0.59 \pm 0.03$ at 450 $\mu$m and $0.36 \pm 0.04$ at 850 $\mu$m, respectively. The obtained slope at 850 $\mu$m is similar to the result in Oph A, 0.33, derived by Pattle et al. (2019). We suggest that the dust grains of NGC 2071IR are well aligned, as in Oph A, with the support of radiation from the central IRS 1, 2, and 3 young stellar objects.

3. We confirm the pinched magnetic field morphology in the central area. This can be explained if the magnetic field is shaped by the rotating disklike structure and the bipolar outflow of IRS 3.

4. We find that the magnetic fields are well aligned with the whole of the IRS 3 bipolar outflow revealed by $^{12}$CO (3−2) and $^{13}$CO (3−2) molecular line emission.

5. We find that the magnetic, turbulent, gravitational, and outflow kinetic energy densities are all comparable to one other in the NGC 2071IR star-forming region. They are $u_B = 2.04(\pm 3.05) \times 10^{-8}$ erg cm$^{-3}$, $u_{\text{turb}} = 3.01(\pm 1.73) \times 10^{-8}$ erg cm$^{-3}$, $u_g = 1.81(\pm 1.51) \times 10^{-8}$ erg cm$^{-3}$, and $u_{\text{outflow}} = 2.33(\pm 0.32) \times 10^{-8}$ erg cm$^{-3}$, respectively.

The James Clerk Maxwell Telescope is operated by the East Asian Observatory on behalf of The National Astronomical Observatory of Japan, Academia Sinica Institute of Astronomy and Astrophysics, the Korea Astronomy and Space Science Institute, and the Center for Astronomical Mega-Science (as well as the National Key R&D Program of China with No. 2017YFA0402700). Additional funding support is provided by the Science and Technology Facilities Council of the United Kingdom and participating universities in the United Kingdom, Canada, and Ireland. Additional funds for the construction of SCUBA-2 were provided by the Canada Foundation for Innovation. D.J. is supported by the National Research Council of Canada and by a Natural Sciences and Engineering Research Council of Canada (NSERC) Discovery Grant. A.S. acknowledges the financial support from NSF through grant AST-1715876. W.K. is supported by the New Faculty Startup Fund from Seoul National University. F.K. and L.F. acknowledge support from the Ministry of Science and Technology of Taiwan, under grant MoST107-2119-M-001-031-MY3 and from Academia Sinica under grant AS-IA-106-M03. C.L.H.H. acknowledges the support of the NAOJ Fellowship and JSPS KAKENHI grants 18K13586 and 20K14527. C.W.L. is supported by the Basic Science Research Program through the National Research Foundation of Korea (NRF) funded by the Ministry of Education, Science and Technology (NRF-2019R1A2C1010851). G.P. is supported by Basic Science Research Program through the National Research Foundation of Korea (NRF) funded by the Ministry of Education (NRF-2020R1A6A3A01100208). K.Q. is partially supported by the National Key R&D Program of China No. 2017YFA0402600, and acknowledges the National Natural Science Foundation of China (NSFC) grant U1731237. The authors wish to recognize and acknowledge the very significant cultural role and reverence that the summit of Maunakea has always had within the indigenous Hawaiian community. We are most fortunate to have the opportunity to conduct observations from this mountain.

## Appendix

We describe how we derive the IRS 3 bipolar outflow kinetic energy density using HARP $^{12}$CO (3−2) and $^{13}$CO (3−2) molecular line data.

First, we estimate the approximate optical depth of the $^{12}$CO (3−2) and $^{13}$CO (3−2) molecular lines from the assumed isotopic line ratio of $X = [^{12}\text{CO}]/[^{13}\text{CO}] = 60$ (Wilson & Rood 1994)

$$R = \frac{T_A^*(^{12}\text{CO})}{T_A^*(^{13}\text{CO})} = \frac{1 - e^{-\tau_{12\text{CO}}}}{1 - e^{-\tau_{13\text{CO}}}} = \frac{1 - e^{-\tau_{12\text{CO}}}}{1 - e^{-\tau_{12\text{CO}}/X}}, \quad (A1)$$

where $T_A^*$ is the corrected antenna temperature for atmospheric attenuation, scattering, and spillover (see Figure A1). This gives $\tau_{12\text{CO}} \sim 4.8$ and $\tau_{13\text{CO}} \sim 0.08$, with a ratio of $R \sim 13$, in the blueshifted outflow, in the velocity range $-0.5 \sim 8.5$ km s$^{-1}$. In the redshifted outflow, the optical depths are $\tau_{12\text{CO}} \sim 3.1$ and $\tau_{13\text{CO}} \sim 0.05$, with a ratio of $R \sim 19$, in the velocity range $10.5 \sim 18.5$ km s$^{-1}$. We do not include the central velocity range of $8.5 \sim 10.5$ km s$^{-1}$ in the optical depth estimation, since $^{12}$CO (3−2) is contaminated by ambient molecular emission.

Second, we derive the column density of $^{13}$CO (3−2) and eventually the total column density of $^{13}$CO, adopting Equations (14.31) and (14.38) derived by Rohlfs & Wilson (2004), since the $^{13}$CO (3−2) line is an optically thin tracer of the bipolar outflow. The adopted equations are

$$N_l = 1.93 \times 10^3 \frac{g_l \nu^2}{g_u A_{ul}} \int T_B dv, \quad (A2)$$

and

$$N(\text{total}) = N(J) \frac{Z}{(2J+1)} \exp\left[\frac{hB_e J(J+1)}{kT}\right], \quad (A3)$$

where $g_l$ and $g_u$ are the statistical weights of the lower and the upper states, respectively; $\nu$ is the frequency in units of GHz; $A_{ul}$ is the Einstein $A$ coefficient in units of s$^{-1}$, which is $2.181 \times 10^{-6}$ s$^{-1}$ for $^{13}$CO (3−2); $Z$ is the partition function; $B_e$ is the rotation constant, which is $5.49 \times 10^{10}$ for $^{13}$CO; $h$ is the Planck constant; $k$ is the Boltzmann constant; and $J$ is the rotational quantum number of the lower state of the transition. We derive an excitation temperature of $T_{\text{ex}} = 13$ K using the $^{12}$CO (3−2) emission, since this line is optically thick. We use Equation (14.33) of Rohlfs & Wilson (2004) to derive the excitation temperature,

$$T_B(\nu) = T_0 \left(\frac{1}{e^{T_0/T_{\text{ex}}} - 1} - \frac{1}{e^{T_0/2.7} - 1}\right)(1 - e^{-\tau_\nu}), \quad (A4)$$

where $T_0 = h\nu/k \approx 16.6$ K at $^{12}$CO (3−2) and $T_B(\nu) \sim 6$ K in the $^{12}$CO (3−2) emission in the outflow.

Third, we estimate the outflow mass. In the blueshifted outflow, we obtain a $^{13}$CO (3−2) column density of $8.29(\pm 1.14) \times 10^{14}$ cm$^{-2}$ using the measured integrated mean brightness temperature $12.1 \pm 1.7$ K km s$^{-1}$ (considering only the calibrator flux error of 13.8%) over an area of $3.1 \times 10^{-7}$ radian$^2$. The derived total $^{13}$CO column density and total column density are $N_{^{13}\text{CO}} = 2.76(\pm 0.38) \times 10^{15}$ cm$^{-2}$ and $N_{\text{total}} = 1.66(\pm 0.23) \times 10^{21}$ cm$^{-2}$, respectively. We adopt abundance ratios of $[^{12}\text{CO}]/[^{13}\text{CO}] = 60$ (Wilson & Rood 1994) and $[H_2]/[^{12}\text{CO}] = 10^4$ for the total column density calculation. The derived mass is then $M_{\text{blue}} = 4.08(\pm 0.56) \times 10^{33}$ g $\approx 2.05 \pm 0.28\ M_\odot$. In the redshifted outflow, we obtain a $^{13}$CO (3−2) column density of $8.49(\pm 1.17) \times 10^{14}$ cm$^{-2}$ using the measured integrated mean brightness temperature $12.4 \pm 1.7$ K km s$^{-1}$ over an area of $3.4 \times 10^{-7}$ radian$^2$. The derived total $^{13}$CO column density and total column density are $N_{^{13}\text{CO}} = 2.83(\pm 0.39) \times 10^{15}$ cm$^{-2}$ and $N_{\text{total}} = 1.70(\pm 0.23) \times 10^{21}$ cm$^{-2}$, respectively. The derived mass is then $M_{\text{red}} = 4.50(\pm 0.62) \times 10^{33}$ g $\approx 2.26 \pm 0.31\ M_\odot$.

Finally, we calculate momentum and energy using the mean velocity derived using the integral $\int T_B v dv / \int T_B dv$ over the area of outflow. The derived mean velocities are $\langle v \rangle \approx 7.1$ km s$^{-1}$ in the blueshifted outflow and $\langle v \rangle \approx 11.2$ km s$^{-1}$ in the redshifted outflow. We then estimate the momentums and energies to be $14.54 \pm 2.01\ M_\odot$ km s$^{-1}$ and $1.03(\pm 0.14) \times 10^{45}$ erg in the blueshifted outflow, and $25.30 \pm 3.50\ M_\odot$ km s$^{-1}$ and $2.82(\pm 0.39) \times 10^{45}$ erg in the redshifted outflow. The estimated energy densities are $1.24(\pm 0.17) \times 10^{-8}$ erg cm$^{-3}$ for the blueshifted outflow, and $3.41(\pm 0.47) \times 10^{-8}$ erg cm$^{-3}$ for the redshifted outflow. We derived a volume of $8.26 \times 10^{52}$ cm$^3$ for both the blue- and redshifted outflows, assuming a conical shape with a radius of 50″ and height of 130″. Therefore, the total kinetic energy and average energy density are $3.85(\pm 0.53) \times 10^{45}$ erg and $2.33(\pm 0.32) \times 10^{-8}$ erg cm$^{-3}$, red respectively (see Table A1).

Table A1
Parameters for the NGC 2071IR Molecular Outflow

| Mass ($M_\odot$) | | | Momentum ($M_\odot$ km s$^{-1}$) | | | Energy ($\times 10^{45}$ erg) | | | Energy Density ($\times 10^{-8}$ erg cm$^{-3}$) |
|---|---|---|---|---|---|---|---|---|---|
| Blue | Red | Total | Blue | Red | Total | Blue | Red | Total | Average |
| $2.05 \pm 0.28$ | $2.26 \pm 0.31$ | $4.31 \pm 0.60$ | $14.54 \pm 2.01$ | $25.30 \pm 3.50$ | $39.85 \pm 5.52$ | $1.03 \pm 0.14$ | $2.82 \pm 0.39$ | $3.85 \pm 0.53$ | $2.33 \pm 0.32$ |

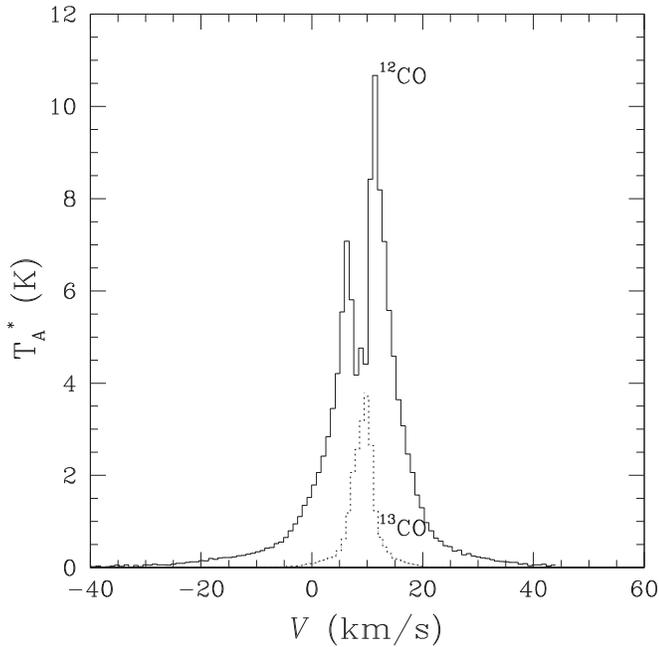

**Figure A1.** Spectra of the $^{12}$CO (3–2) and $^{13}$CO (3–2) molecular line emission toward the NGC 2071IR outflow region.

**ORCID iDs**

A-Ran Lyo ⓘ https://orcid.org/0000-0002-9907-8427
Jongsoo Kim ⓘ https://orcid.org/0000-0002-1229-0426
Doug Johnstone ⓘ https://orcid.org/0000-0002-6773-459X
Kate Pattle ⓘ https://orcid.org/0000-0002-8557-3582
Woojin Kwon ⓘ https://orcid.org/0000-0003-4022-4132
Pierre Bastien ⓘ https://orcid.org/0000-0002-0794-3859
Takashi Onaka ⓘ https://orcid.org/0000-0002-8234-6747
Ji-Hyun Kang ⓘ https://orcid.org/0000-0001-7379-6263
Ray Furuya ⓘ https://orcid.org/0000-0003-0646-8782
Charles L. H. Hull ⓘ https://orcid.org/0000-0002-8975-7573
Motohide Tamura ⓘ https://orcid.org/0000-0002-6510-0681
Patrick M. Koch ⓘ https://orcid.org/0000-0003-2777-5861
Derek Ward-Thompson ⓘ https://orcid.org/0000-0003-1140-2761
Thiem Hoang ⓘ https://orcid.org/0000-0003-2017-0982
Doris Arzoumanian ⓘ https://orcid.org/0000-0002-1959-7201
Chin-Fei Lee ⓘ https://orcid.org/0000-0002-3024-5864
Do-Young Byun ⓘ https://orcid.org/0000-0003-1157-4109
Florian Kirchschlager ⓘ https://orcid.org/0000-0002-3036-0184
Yasuo Doi ⓘ https://orcid.org/0000-0001-8746-6548
Kee-Tae Kim ⓘ https://orcid.org/0000-0003-2412-7092
Jihye Hwang ⓘ https://orcid.org/0000-0001-7866-2686
Sang-Sung Lee ⓘ https://orcid.org/0000-0002-6269-594X
Geumsook Park ⓘ https://orcid.org/0000-0001-8467-3736
Hyunju Yoo ⓘ https://orcid.org/0000-0002-8578-1728
Eun Jung Chung ⓘ https://orcid.org/0000-0003-0014-1527
Steve Mairs ⓘ https://orcid.org/0000-0002-6956-0730
Archana Soam ⓘ https://orcid.org/0000-0002-6386-2906
Tie Liu ⓘ https://orcid.org/0000-0002-5286-2564
Xindi Tang ⓘ https://orcid.org/0000-0002-4154-4309
Simon Coudé ⓘ https://orcid.org/0000-0002-0859-0805
Philippe André ⓘ https://orcid.org/0000-0002-3413-2293
Tyler L. Bourke ⓘ https://orcid.org/0000-0001-7491-0048
Zhiwei Chen ⓘ https://orcid.org/0000-0003-0849-0692
Wen Ping Chen ⓘ https://orcid.org/0000-0002-5519-0628
Tao-Chung Ching ⓘ https://orcid.org/0000-0001-8516-2532
Jungyeon Cho ⓘ https://orcid.org/0000-0003-1725-4376
Antonio Chrysostomou ⓘ https://orcid.org/0000-0002-9583-8644
David Eden ⓘ https://orcid.org/0000-0002-5881-3229
Erica Franzmann ⓘ https://orcid.org/0000-0003-2142-0357
Per Friberg ⓘ https://orcid.org/0000-0002-8010-8454
Rachel Friesen ⓘ https://orcid.org/0000-0001-7594-8128
Gary Fuller ⓘ https://orcid.org/0000-0001-8509-1818
Tim Gledhill ⓘ https://orcid.org/0000-0002-2859-4600
Sarah Graves ⓘ https://orcid.org/0000-0001-9361-5781
Qilao Gu ⓘ https://orcid.org/0000-0002-2826-1902
Jannifer Hatchell ⓘ https://orcid.org/0000-0002-4870-2760
Martin Houde ⓘ https://orcid.org/0000-0003-4420-8674
Tsuyoshi Inoue ⓘ https://orcid.org/0000-0002-7935-8771
Shu-ichiro Inutsuka ⓘ https://orcid.org/0000-0003-4366-6518
Kazunari Iwasaki ⓘ https://orcid.org/0000-0002-2707-7548
Miju Kang ⓘ https://orcid.org/0000-0002-5016-050X
Akimasa Kataoka ⓘ https://orcid.org/0000-0003-4562-4119
Koji Kawabata ⓘ https://orcid.org/0000-0001-6099-9539
Francisca Kemper ⓘ https://orcid.org/0000-0003-2743-8240
Gwanjeong Kim ⓘ https://orcid.org/0000-0003-2011-8172
Mi-Ryang Kim ⓘ https://orcid.org/0000-0002-1408-7747
Shinyoung Kim ⓘ https://orcid.org/0000-0001-9333-5608
Jason Kirk ⓘ https://orcid.org/0000-0002-4552-7477
Masato I. N. Kobayashi ⓘ https://orcid.org/0000-0003-3990-1204
Vera Könyves ⓘ https://orcid.org/0000-0002-3746-1498
Takayoshi Kusune ⓘ https://orcid.org/0000-0002-9218-9319
Jungmi Kwon ⓘ https://orcid.org/0000-0003-2815-7774
Kevin Lacaille ⓘ https://orcid.org/0000-0001-9870-5663
Shih-Ping Lai ⓘ https://orcid.org/0000-0001-5522-486X
Chi-Yan Law ⓘ https://orcid.org/0000-0003-1964-970X
Jeong-Eun Lee ⓘ https://orcid.org/0000-0003-3119-2087
Yong-Hee Lee ⓘ https://orcid.org/0000-0001-6047-701X
Hyeseung Lee ⓘ https://orcid.org/0000-0003-3465-3213
Di Li ⓘ https://orcid.org/0000-0003-3010-7661
Hua-Bai Li ⓘ https://orcid.org/0000-0003-2641-9240
Junhao Liu ⓘ https://orcid.org/0000-0002-4774-2998
Sheng-Yuan Liu ⓘ https://orcid.org/0000-0003-4603-7119
Xing Lu ⓘ https://orcid.org/0000-0003-2619-9305
Masafumi Matsumura ⓘ https://orcid.org/0000-0002-6906-0103
Brenda Matthews ⓘ https://orcid.org/0000-0003-3017-9577
Gerald Moriarty-Schieven ⓘ https://orcid.org/0000-0002-0393-7822

Tetsuya Nagata https://orcid.org/0000-0001-9264-9015
Fumitaka Nakamura https://orcid.org/0000-0001-5431-2294
Nagayoshi Ohashi https://orcid.org/0000-0003-0998-5064
Harriet Parsons https://orcid.org/0000-0002-6327-3423
Tae-soo Pyo https://orcid.org/0000-0002-3273-0804
Lei Qian https://orcid.org/0000-0003-0597-0957
Keping Qiu https://orcid.org/0000-0002-5093-5088
Ramprasad Rao https://orcid.org/0000-0002-1407-7944
Jonathan Rawlings https://orcid.org/0000-0001-5560-1303
Mark G. Rawlings https://orcid.org/0000-0002-6529-202X
John Richer https://orcid.org/0000-0002-9693-6860
Andrew Rigby https://orcid.org/0000-0002-3351-2200
Giorgio Savini https://orcid.org/0000-0003-4449-9416
Yoshito Shimajiri https://orcid.org/0000-0001-9368-3143
Hiroko Shinnaga https://orcid.org/0000-0001-9407-6775
Ya-Wen Tang https://orcid.org/0000-0002-0675-276X
Kohji Tomisaka https://orcid.org/0000-0003-2726-0892
Le Ngoc Tram https://orcid.org/0000-0002-6488-8227
Serena Viti https://orcid.org/0000-0001-8504-8844
Jia-Wei Wang https://orcid.org/0000-0002-6668-974X
Hongchi Wang https://orcid.org/0000-0003-0746-7968
Hsi-Wei Yen https://orcid.org/0000-0003-1412-893X
Hyeong-Sik Yun https://orcid.org/0000-0001-6842-1555
Chuan-Peng Zhang https://orcid.org/0000-0002-4428-3183